\documentclass[amsmath,amssymb,superscriptaddress,aps,pra,twocolumn,showpacs,reprint]{revtex4-1}
\pdfoutput=1

\usepackage{amsmath}
\usepackage{bm}
\usepackage{graphicx}

\usepackage{multirow}
\usepackage{color}

\usepackage[USenglish]{babel}

\renewcommand {\vec} [1] {{\bm #1}}
\newcommand {\s} [1] {{\(_{#1}\)}}
\newcommand {\review}[1]{#1}
\usepackage[colorlinks=true,citecolor=blue,urlcolor=blue,linkcolor=blue]{hyperref}

\begin{document}

\title[Real-space DFT on GPUs]{Real-space density functional theory on
  graphical processing units: computational approach and comparison to
  Gaussian basis set methods}

\author{Xavier Andrade} 
\author{Al\'an Aspuru-Guzik}
\email{aspuru@chemistry.harvard.edu}
\affiliation{Department of Chemistry and Chemical Biology, Harvard
  University, 12 Oxford Street, Cambridge, MA 02138, United States}


\begin{abstract}
  We discuss the application of graphical processing units (GPUs) to
  accelerate real-space density functional theory (DFT) calculations.
  To make our implementation efficient, we have developed a scheme to
  expose the data parallelism available in the DFT approach; this is
  applied to the different procedures required for a real-space DFT
  calculation.  We present results for current-generation GPUs from
  AMD and Nvidia, which show that our scheme, implemented in the free
  code \textsc{octopus}, can reach a sustained performance of up to 90
  GFlops for a single GPU, representing \review{a} significant speed-up when
  compared to the CPU version of the code.  Moreover, for some systems
  our implementation can outperform a GPU Gaussian basis set code,
  showing that the real-space approach is a competitive alternative
  for DFT simulations on GPUs.
\end{abstract}

\maketitle

\section{Introduction}

For many years, the constant reduction in the size of the transistors,
as described by Moore's law~\cite{Moore1965}, has been translated into
an increment of the processing capacity of central processing units
(CPUs). 
%
However, due to the limitations in efficiency and power consumption
\review{related to the breakdown of Dennard
  scaling~\cite{Dennard1974,*Bohr2009}}, CPU designers moved towards
parallel processing to profit from the increasing number of
transistors. This trend towards parallelism can be seen in current
CPUs, that have multiple cores, with each core capable of executing
multiple threads and containing vectorial processing units that
operate simultaneously on several sets of values.

Simultaneously, a more parallel kind of processor appeared: the
graphical processing unit (GPU). Originally designed for real-time
rendering of images, a computationally intensive and highly-parallel
task, modern GPUs are also suitable for general purpose computing, in
particular for high-performance numerical simulations. They typically
have thousands of execution units, that give them approximately one
order of magnitude higher processing power than a CPU. This difference
is explained by different design strategies: while a single
instruction may be executed faster on a CPUs, GPUs can execute
thousands of them in parallel.

In the last years there has been a considerable interest in applying
GPUs to computational science. While in some areas of atomistic
simulations GPUs are becoming a standard tool~\cite{Harju2013}, in the
electronic structure domain, and in particular in density functional
theory (DFT)~\cite{Hohenberg1964,*Kohn1965}, the adoption of GPUs has
been slower. The first full electronic-structure implementation on
GPUs was \textsc{terachem}, presented by Ufimtsev and Mart\'inez in
2008~\cite{Ufimtsev2008a}. Currently, several electronic structure
codes have also incorporated some degree of GPU
acceleration~\cite{Yasuda2008, Vogt2008, Genovese2009, Watson2010,
  Tomono2010, Andrade2012c, Andrade2012a, Maintz2011, DePrince2011,
  Spiga2012, Maia2012, Hacene2012, Esler2012,
  Hakal2013,Jia2013a,*Jia2013b, Hutter2013}.

Still, how to get the most out of a GPU for modeling electronic
systems is an active area of research~\cite{Spiga2012,BhaskaranNair2013,Titov2013}, as
simulation approaches that are efficient on a CPU might not be as
efficient on a GPU. These approaches can be improved or replaced by
other methods that are better suited to massively parallel
architectures. In this respect, the large diversity of methods used
for electronic structure methods by chemists and physicist, offers an
interesting starting point to explore the application of GPUs to the
simulation of electronic systems.

In this work we focus on one particular approach for electronic
structure, real-space DFT, and how it can be adapted to GPUs. While
not as widely used as basis-set methods, the real-space grid
discretization is a popular alternative for DFT
simulations~\cite{Becke1989, Chelikowsky1994, Briggs1995, Fattebert2000,
  Fattebert2003, Beck2000, Marques2003, Torsti2003, Hirose2005,
  Mortensen2005, Kronik2006, Yabana2006, Hernandez2007, Iwata2010}.  Its main
features are the flexibility to model different types of electronic
systems, the systematic control of the discretization error, and its
potential for parallelization in distributed memory systems with
thousands of
processors~\cite{Andrade2012a,Bernholc2008,Enkovaara2010,Hasegawa2011}.

The development of an efficient GPU implementation does not only
involve rewriting and optimizing low-level routines for the GPU. For
complex scientific software, choosing an appropriate design strategy
for the entirety of the code can be fundamental for optimal GPU
performance. This work is mainly focused on this issue: we have
developed a scheme to apply DFT efficiently on GPUs by exposing the
available parallelism to the low-level routines.

Our approach was developed for the implementation of GPU support in
the \textsc{octopus} code~\cite{Marques2003,Castro2006,Andrade2012a}
and is freely available under an open source
license~\cite{octopusdownload}. \textsc{Octopus} is used by several
research groups for theoretical development~\cite{Burnus2005,
  Botti2008, Andrade2009, Rasanen2010, Helbig2011, DeGiovannini2012,
  Elliot2012, Andrade2012b} and applications in different fields of
chemistry and physics~\cite{Wasserman2008, Malloci2008, Botti2009,
  Vila2010, Zhang2011, Bonaca2011, AvendanoFranco2012, Castro2013,
  Rozzi2013, Rasanen2013}. In this article, we describe in detail our
general strategy and its application to the different procedures
required for real-space DFT, extending previous results for real-time
time-dependent DFT~\cite{Andrade2012c,Andrade2012a}. Our GPU
implementation is based on OpenCL~\cite{Munshi2010}, a standard and
portable framework for writing code for parallel processors, so it can
run on GPUs, CPUs, and other processing devices, from different
vendors.

In order to assess the efficiency of our implementation, we perform a
series of tests involving top-end GPUs from Nvidia and AMD, and a set
of molecular systems of different sizes. We provide different
indicators that illustrate the performance of our implementation:
numerical throughput (number of floating point operations executed per
unit of time), total calculation times, and comparisons with the CPU
version of the code and a different GPU-DFT implementation. These
results show that real-space DFT is an interesting and competitive
approach for GPU-accelerated electronic structure calculations.


\section{Real-space density functional theory}
\label{sec:rsdft}

In the Kohn-Sham (KS) formulation of DFT, the electronic density of an
interacting electronic system, \(n(\vec{r})\), is generated by a set
of single-particle orbitals, or states, \(\varphi_k(\vec{r})\). These
orbitals are generated by the KS equations~\cite{Hohenberg1964,*Kohn1965}
\begin{subequations}
\label{eq:ks}
\begin{align}
  \label{eq:ksorb}
  H[n]\varphi_k(\vec{r}) & = \epsilon_k\,\varphi_k(\vec{r}) \\
  \label{eq:ksden}
  n(\vec{r}) &= \sum_{k=1}^{N}\varphi_k^*(\vec{r})\varphi_k(\vec{r})\ ,
\end{align}
\end{subequations}
where the \(H\) operator is the KS effective single-particle
Hamiltonian, (atomic units are used throughout)
\begin{equation}
  \label{eq:kshamiltonian}
  H[n] = -\frac12\nabla^2 + v_{\mathrm{ext}}(\vec{r}) + v_{\mathrm{hxc}}[n](\vec{r},t)\ .
\end{equation}
The external potential \(V_{\mathrm{ext}}\) contains the nuclear
potential and other external fields that may be present,
\(V_{\mathrm{hxc}}\) represents the electron-electron interaction and
is usually divided in the Hartree term, that contains the classical
electrostatic interaction between electrons, and the exchange and
correlation (XC) potential.

To solve the KS equations numerically, the orbitals, the density, and
other fields need to be represented as a finite set of numbers. The
selection of the discretization scheme is probably the most important
aspect in the numerical solution of the electronic structure
problem. Traditionally, a basis set expansion is used: atomic orbitals
for molecules, and plane waves for crystalline systems. In the
real-space approach, instead of a basis, fields are discretized in a
grid. This provides a simple and flexible scheme that is suitable to
model both finite and periodic systems~\cite{Natan2008}. The electron ion interaction
is modeled by the pseudo-potential approximation, or the
projector-augmented-wave method~\cite{Mortensen2005}, that remove the
problem of representing the hard Coulomb potential, so uniform grids
can be used.

\begin{figure}
\includegraphics[width=\columnwidth]{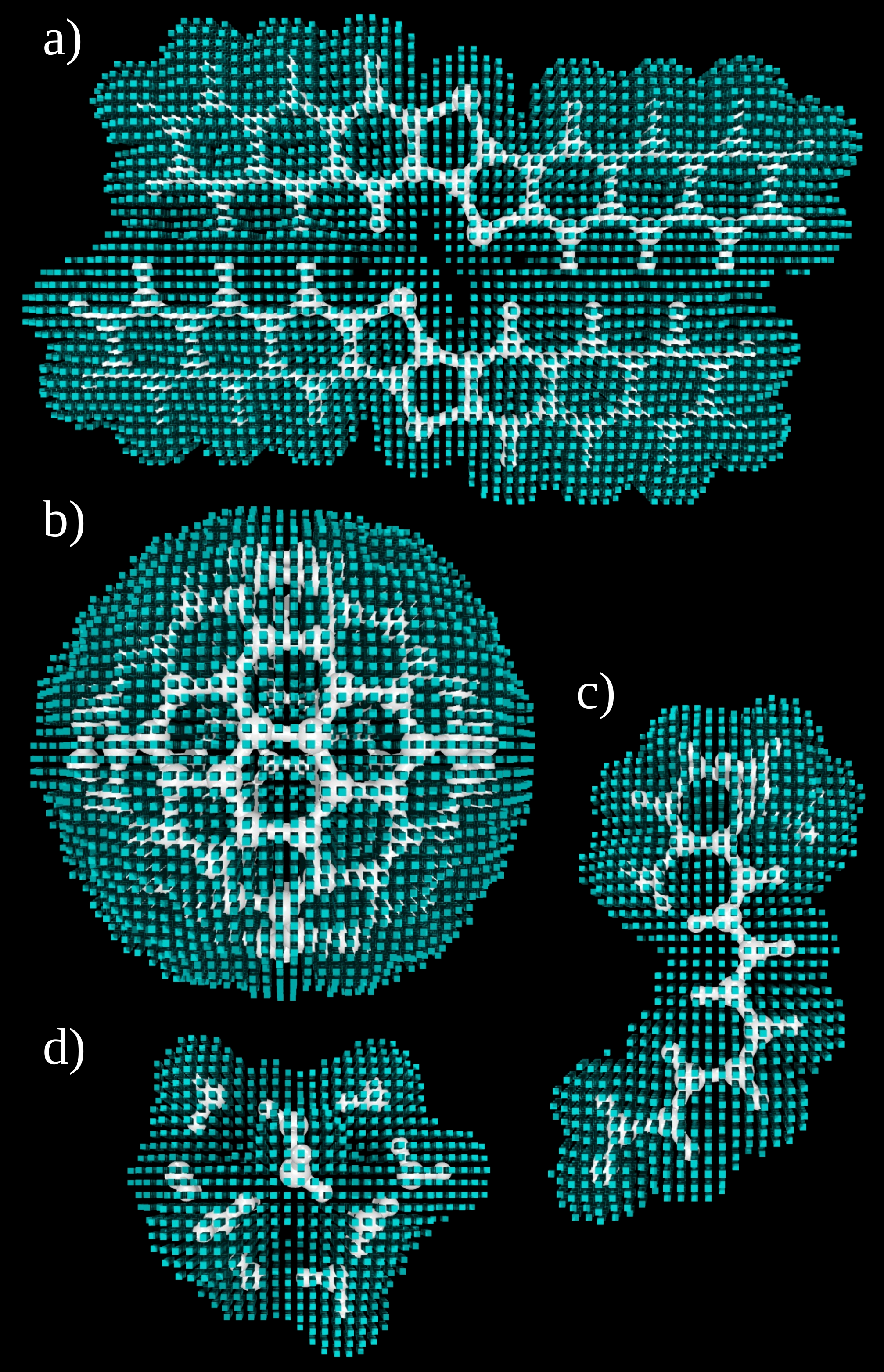}
\caption{Examples of real-space grids adapted to the shape of
  different molecular systems: a) DAT-thiophane dimer b) C\s{180}
  fullerene c) cis-retinal d) water cluster. The (cyan) cubes mark the
  position of the grid points. For visualization purposes, we
  represent smaller and coarser grids than the ones used for actual
  calculations.
  \label{fig:struct}}
\end{figure}

One of the main advantages of the real-space grid approach is that the
discretization error can be controlled systematically by reducing the
spacing and increasing the size of the box. Of course, this increases
the number of points and, proportionally, the time and memory cost of
the calculation. To keep the number of grid points to a minimum, our
implementation uses arbitrarily-shaped grids. This choice makes the
code more complex but allows for an important reduction in grid size
compared with a simpler cubic grid. For molecular systems we use a
\review{uniform grid whose shape is given by the union of spheres
  around each atom, as shown in Fig.~\ref{fig:struct}. This strategy
  avoids placing points in regions where the value of the density is
  not significant for the desired accuracy.}

\section{Numerical solution of real-space density functional theory}
\label{sec:procs}

\begin{figure}
\includegraphics[width=\columnwidth]{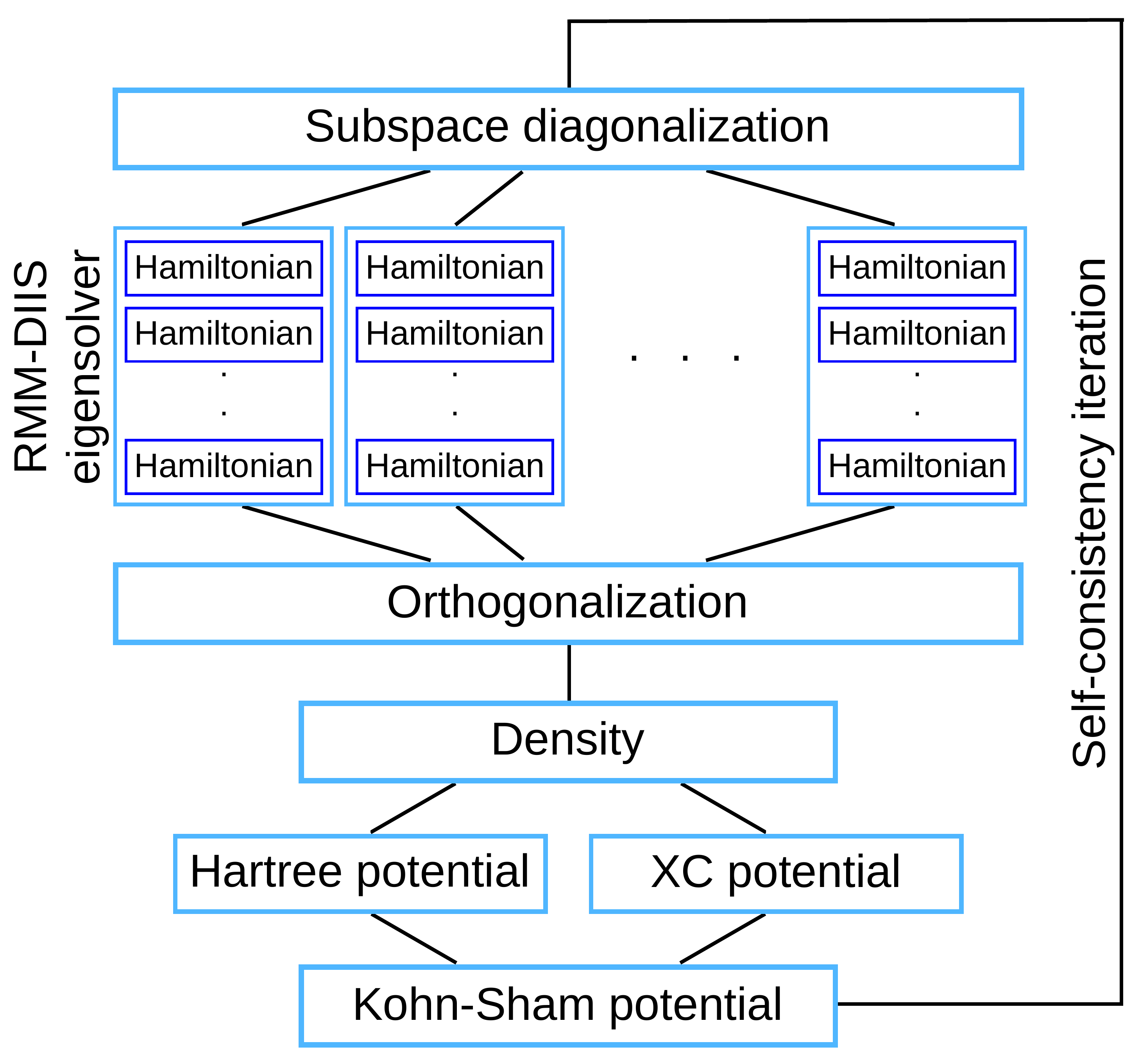}
\caption{Schematic of a density functional theory calculation in
  real-space using a self-consistency scheme and the residual
  minimization-direct inversion in the iterative subspace (RMM-DIIS)
  eigensolver. The boxes represent the different numerical procedures
  that need to be performed. \label{fig:groundstate}}
\end{figure}

We now describe the numerical procedure to solve the KS equations in
real-space. As it is standard in Hartree-Fock (HF) and DFT, in order
to account for the \review{nonlinearity} introduced by the density dependence
in eq.~(\ref{eq:ks}) a self-consistent field (SCF) iterative scheme is
used. A new set of orbitals and density are generated each SCF
iteration; this involves several numerical procedures, that are shown
in Fig.~\ref{fig:groundstate}.

Every SCF step we need to find the lower eigenvectors and eigenvalues
of the KS Hamiltonian for a fixed density. In real-space, the
discretization of the KS Hamiltonian, eq.~(\ref{eq:kshamiltonian}), is
done using a high-order finite differences
representation~\cite{Chelikowsky1994}. As this results in a sparse
operator, the diagonalization is done using iterative methods that do
not require the Hamiltonian matrix to be built explicitly, only to be
applied as an operator. In this work, we use the efficient residual
minimization-direct inversion in the iterative subspace (RMM-DIIS)
eigensolver~\cite{Wood1985,Kresse1996} (not to be confused with the
DIIS SCF scheme~\cite{Pulay1980}). To precondition the eigensolver, we
use the filter operator proposed by Saad \textit{et al.}~\cite{Saad1996}.

In practice, it is not worth it to find a converged solution of the
eigenvalue problem at each SCF iteration: instead we do a fixed number
of eigensolver iterations per step. In this manner, the eigenproblem
convergence is achieved towards the end of the SCF cycle.

The RMM-DIIS scheme requires the application of the KS Hamiltonian and
two additional procedures that act over the whole set of orbitals:
orthogonalization and subspace diagonalization. Given a set of
orbitals, the orthogonalization procedure performs a linear
transformation that generates a new orthogonal set. Similarly,
subspace diagonalization is an effective method to remove
contamination between orbitals. It calculates the representation of
the KS Hamiltoninan in the subspace spanned by a set of orbitals, and
generates a new set where the subspace Hamiltonian is diagonal.

After the eigensolver steps and the posterior orthogonalization, a new
set of orbitals and a new density are obtained; this density is mixed
with the densities from previous steps to generate a new guess density
according to the Broyden scheme~\cite{Broyden1965,Srivastava1984}.

From the new guess density, the new KS effective potential is
calculated. Numerically, the most expensive part of this step is
obtaining the Hartree potential, \(V_\mathrm{H}\), that requires the
solution of the Poisson equation
\begin{equation}
\nabla^2V_\mathrm{H}(\vec{r})=-4\pi\,n(\vec{r})\ .\label{eq:poisson}
\end{equation}
In our implementation, we use a Poisson solver based on fast Fourier
transforms. The XC potential, \(v_\mathrm{xc}\), also needs to be
recalculated. This is approximated by a local or semi-local expression
that is evaluated directly on the grid.

\section{General GPU optimization strategy}
\label{sec:gpu_general}

In this section we discuss the general scheme that we have developed
to solve efficiently the real-space DFT equation on GPUs. This
strategy was designed taking into account the strengths and weaknesses
of the current generation of GPUs, but is also effective for CPUs with
vectorial floating point units.

For optimum efficiency, GPUs need to operate simultaneously over large
amounts of data, so that the numerous independent operations fill the
execution units and hide operation and memory latency (the time it
takes the result of an instruction to be available to other
instructions). A way to fulfill this requirement is to expose
data-parallelism to the low-level routines. For example, if the same
operation needs to be performed over certain data objects, the
routines should receive as an argument a group of those objects,
instead of operating over one object per call.

\begin{figure}
  \centering
  \includegraphics[width=\columnwidth]{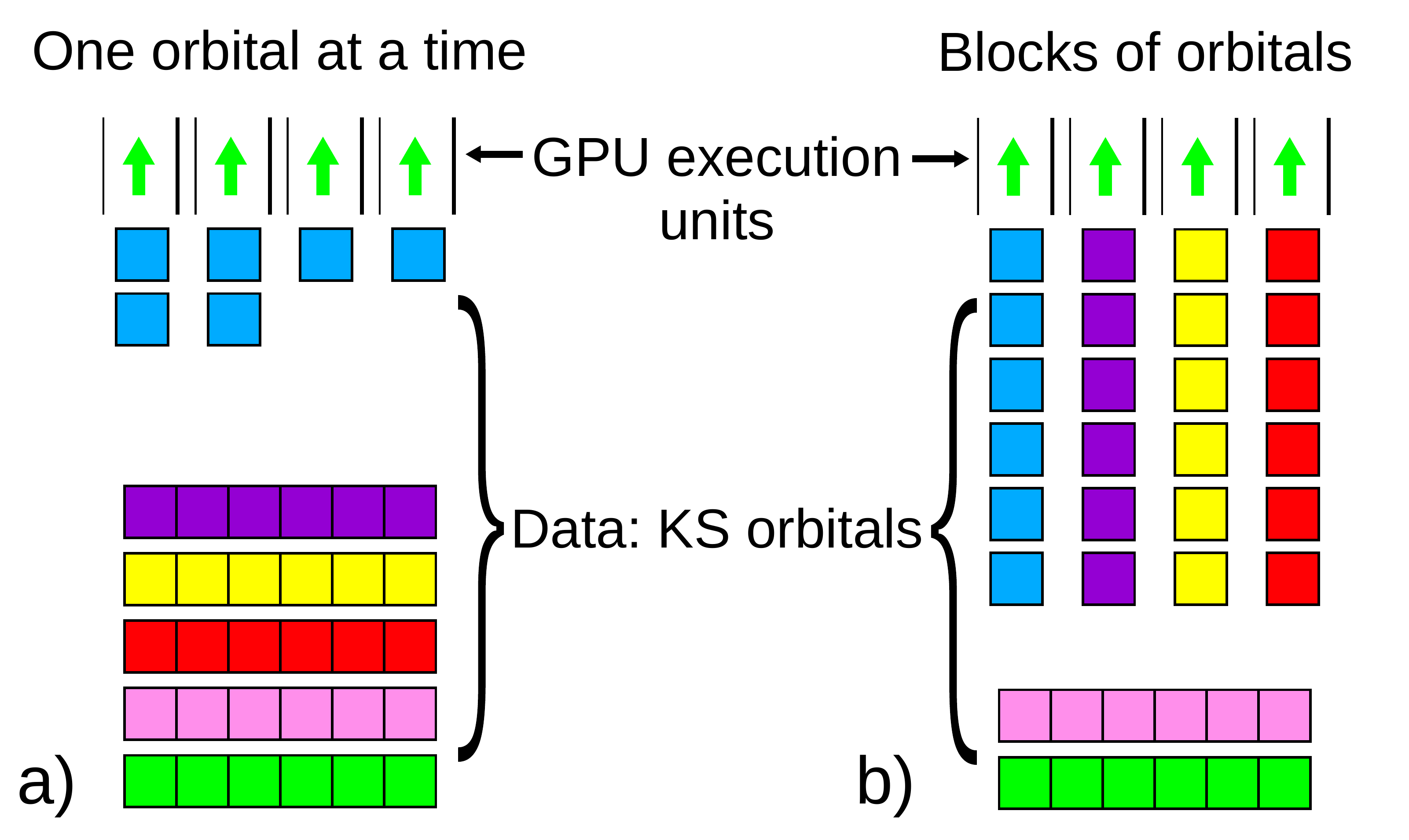}
  \caption{Scheme illustrating the blocks of orbitals strategy for DFT
    on GPUs. a) Operating on a single orbital might not provide enough
    parallelism for the GPU to perform efficiently. b) By operating
    simultaneously over several orbitals there is a larger degree of
    data parallelism and there is less divergence among GPU threads.}
  \label{fig:blocks}
\end{figure}

In order to expose parallelism in the DFT case, our GPU optimization
strategy is based on the concept of blocks of KS orbitals. Instead of
acting over a single KS orbital at a time, performance critical
routines receive a group of orbitals as argument. By operating
simultaneously over several orbitals, the amount of parallelism
exposed to the processor is increased considerably. In
Fig.~\ref{fig:blocks} we show a scheme of how this concept works.

The blocks-of-orbitals strategy has an additional advantage: in a GPU,
threads are divided in groups of 32 (Nvidia) or 64 (AMD), called warps
or wavefronts; for efficient execution all threads in a warp must
execute exactly the same instruction sequence. Since the same
operation has to be performed over each orbital, we can assign
operations corresponding to different orbitals to different threads in
a warp. This ensures that the execution within each warp is regular,
without divergences in the instruction sequence. In a CPU, vectorial
floating point units play a similar role as warps.

A possible drawback of the block-of-orbitals approach is that memory
access issues might appear, as working with larger amount of data can
saturate caches and reduce their ability to speed-up memory
access. This is especially true for CPUs, which rely more on caches
than GPUs. Larger blocks can also increase the amount of memory
required for temporary data. So using blocks that are too large might
be detrimental for performance.

In our implementation the number of orbitals in a block, or
block-size, is variable and controlled at execution time. Ideally the
block-size should be an integer multiple of the warp-size. This might
not be possible if not enough orbitals are available, in such a case
the block-size should be a divisor of the warp size. Following these
considerations we restrict our block-size to be a small power of
two~\footnote{This has the additional advantage that the integer
  multiplication by the block-size, required for array address
  calculations, can be done using the cheaper bit-shift
  instructions.}.

\begin{figure}[t]
  \centering
  \includegraphics[width=1.0\columnwidth]{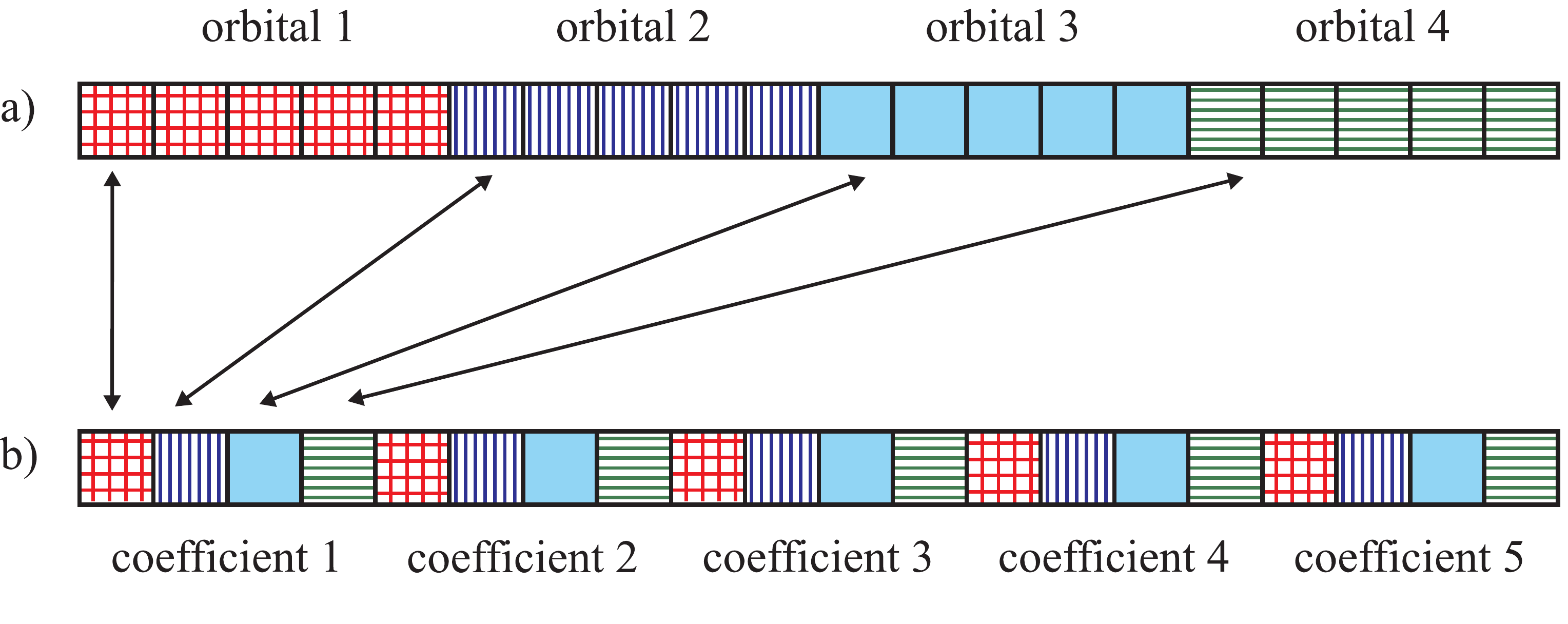}
  \caption{Example of memory layout for a block of 4 orbitals with 5
    coefficients each: (a)~Standard memory layout where each orbital
    is contiguous in memory. (b)~Optimal memory layout where all the
    coefficients in a block are contiguous. The arrows indicate the
    relation of the position of the first coefficient in both
    schemes.}
  \label{fig:memory_layout}
\end{figure}

The way blocks of orbitals are stored in memory is also fundamental
for optimal performance. A natural scheme would be to store the
coefficients for each orbital (Fig.~\ref{fig:memory_layout}a)
contiguously in memory, so that each orbital in a block can be easily
accessed. However, memory access is usually more efficient when
threads access adjacent memory locations as loads or stores go to the
same cache-lines. Since in our approach consecutive threads are
assigned to different orbitals, we order blocks by the orbital index
first and then by the discretized \(\vec{r}\)-index, ensuring that
adjacent threads will access adjacent memory locations
(Fig.~\ref{fig:memory_layout}b).

\begin{figure}[t]
  \centering
  \includegraphics[width=1.0\columnwidth]{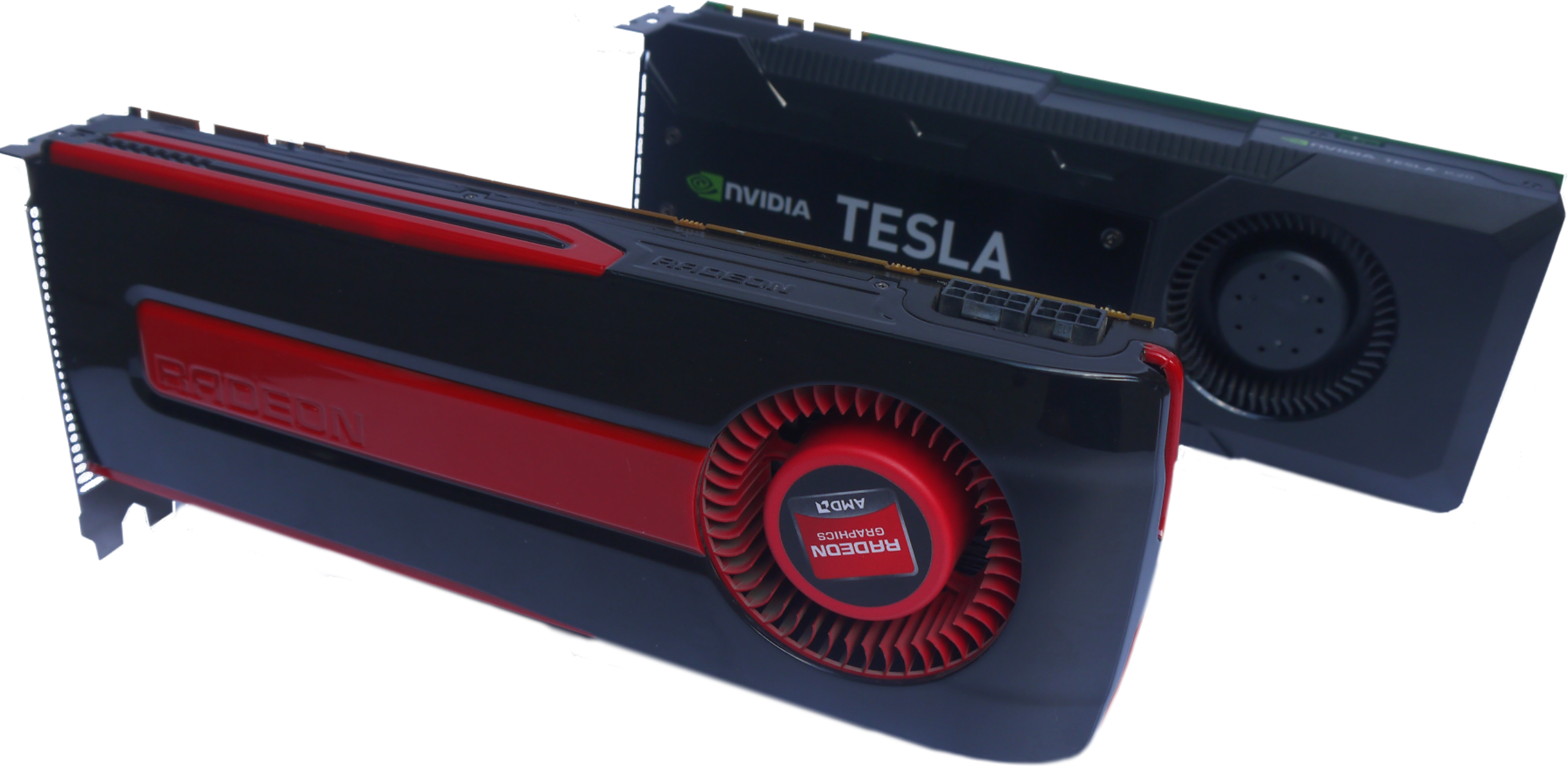}
  \caption{AMD Radeon HD 7970 and a Nvidia Tesla K20 GPU cards used
    for the numerical tests.}
  \label{fig:gpus}
\end{figure}

In the following sections, we show how these general strategies are
applied to the different numerical procedures that were introduced in
section~\ref{sec:procs}. For each operation we show the numerical
performance that our implementation obtains for a test system,
\(\beta\)-cyclodextrin, on an Intel Core i7 3820 CPU and two GPUs, an AMD
Radeon HD 7970 and a Nvidia Tesla K20 (shown in
Fig.~\ref{fig:gpus}). Details about the platforms and the calculations
can be found in section~\ref{sec:methods}.

\section{Kohn-Sham Hamiltonian}
\label{sec:hamiltonian}

The application of the KS Hamiltonian, eq.~(\ref{eq:kshamiltonian}),
is the basic operation of the real-space DFT approach, as such, it is
the first target for efficient GPU execution. Moreover, the KS
Hamiltonian application is also used in other DFT-based simulations
like on-the-fly molecular dynamics~\cite{Tuckerman1994,*Alonso2008},
and response calculations in time~\cite{Yabana1996,*Castro2004} and
frequency domains~\cite{Baroni2001,*Andrade2007}.

As a matrix, the real-space KS Hamiltonian operator is sparse, with a
number of coefficients that is proportional to the number of grid
points. While the matrix could be stored in a sparse form, it is not
convenient to do so. It is more efficient to use it in operator form,
with three different terms that are applied independently: the kinetic
energy operator, the local potential, and the non-local potential.

\subsection{Kinetic energy operator}

\begin{figure}
  \centering
  \includegraphics*[width=0.9\columnwidth]{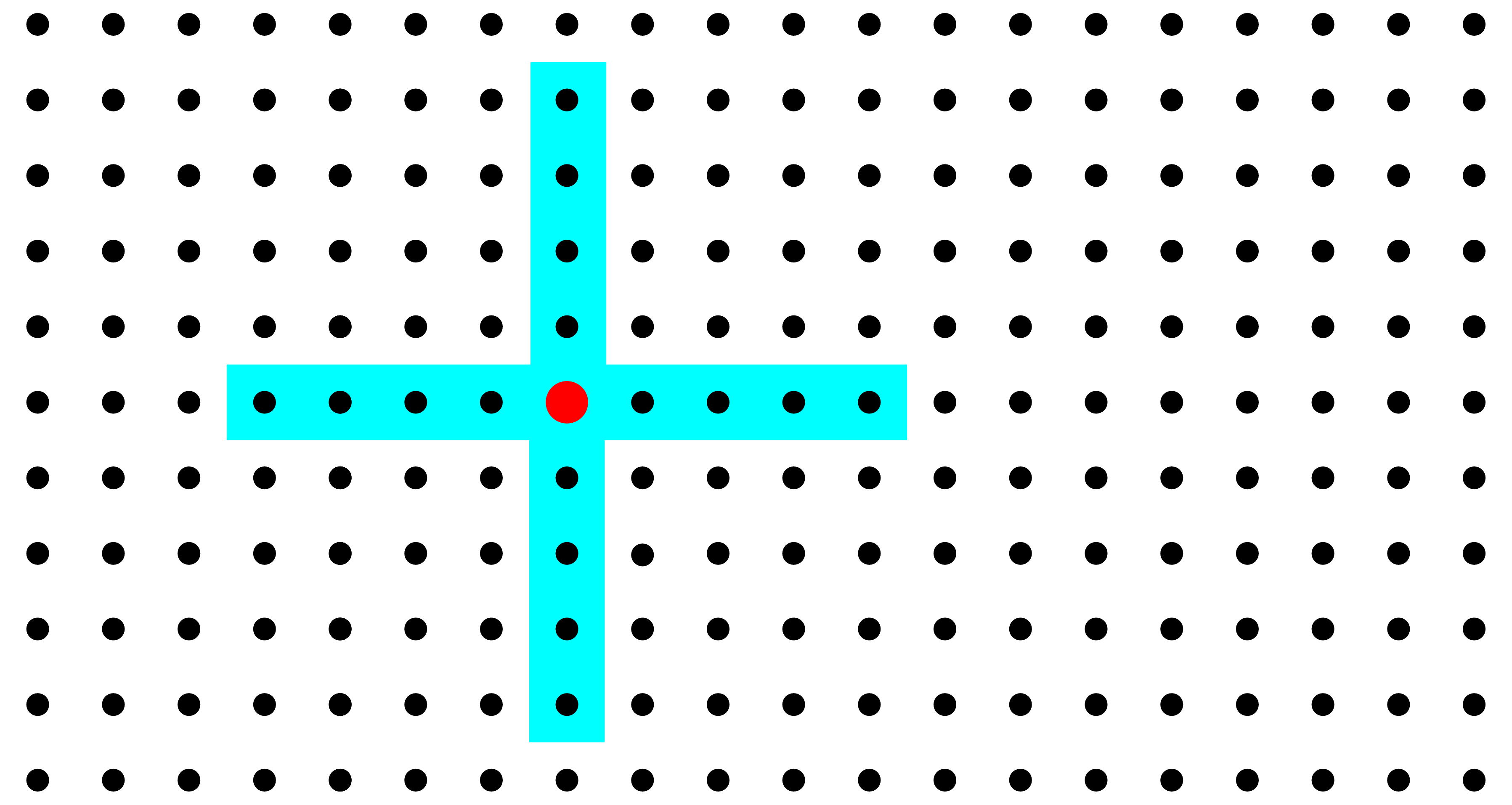}
  \caption{Example of a stencil for the fourth-order Laplacian in a 2D
    grid. The values from points in the colored region are used to
    calculate the Laplacian in the central (red) point.}
  \label{fig:stencil}
\end{figure}

In real-space the kinetic-energy operator corresponds to the Laplacian
differential operator. While in a basis-set approach this term is
calculated exactly, in real-space the Laplacian is approximated using
high-order finite-differences~\cite{Chelikowsky1994}.  Numerically,
this is a stencil calculation, where the value at each point is
calculated as a linear combination of the neighboring-point
values. The stencil represents the grid points used in the calculation
of the differential operator, see Fig.~\ref{fig:stencil} for an
example.
In the
simulations presented in this paper we use a fourth-order
approximation that in 3D results in a stencil size of 25.

\review{Since stencil calculations are common in scientific and
  engineering applications, their optimization in CPU and GPU
  architectures has received considerable
  interest~\cite{Peng2009,Datta2009,Dursun2009,Treibig2011,delaCruz2011,Henretty2011,Holewinski2012}. In
  our approach the stencil is applied over several orbitals at once,
  avoiding some of the performance issues that appear in the
  application of a stencil to a single dataset, in particular with
  respect to vectorization~\cite{Henretty2011}.  }

On the GPU, to perform the application of the Laplacian over a block
of orbitals the threads are arranged in a two-dimensional grid: the
first dimension corresponds to the orbital index and the second to the
point index. The first task of each group of threads is to find the
location of the neighboring points in the input array. Since the grid
has an arbitrary shape, this location cannot be easily calculated and
we need to use a table of neighbors~\footnote{To avoid working with a
  full table of neighbors, we use a compact form~\cite{Andrade2010}.}. Once the neighbor addresses
are obtained, each thread iterates over the stencil position loading
the neighbor value, multiplying it by the corresponding weight and
accumulating the result.



Memory access is usually the limiting factor for the performance of
the finite-difference operators\review{~\cite{Datta2009}}, as for each point we
need to iterate over the stencil loading values that are only used for
one multiplication and one addition. As the values of the neighbors
are scattered, memory access is not regular. This part of the problem
is addressed by using blocks of orbitals: since the Laplacian is
calculated over a group of orbitals at a time, for each point of the
stencil we load several values, one per orbital in the block, that are
contiguous in memory. This makes memory access more regular and hence
more efficient for both GPUs and CPUs.

Still, a potential problem with memory access persists. As each input
value of the stencil has to be loaded several times, ideally it should
be loaded from main memory once and kept in cache for subsequent
uses. Unfortunately, as the stencil operation has poor memory
locality, this is not always the case.

\begin{figure}
  \centering
  \includegraphics[width=\columnwidth]{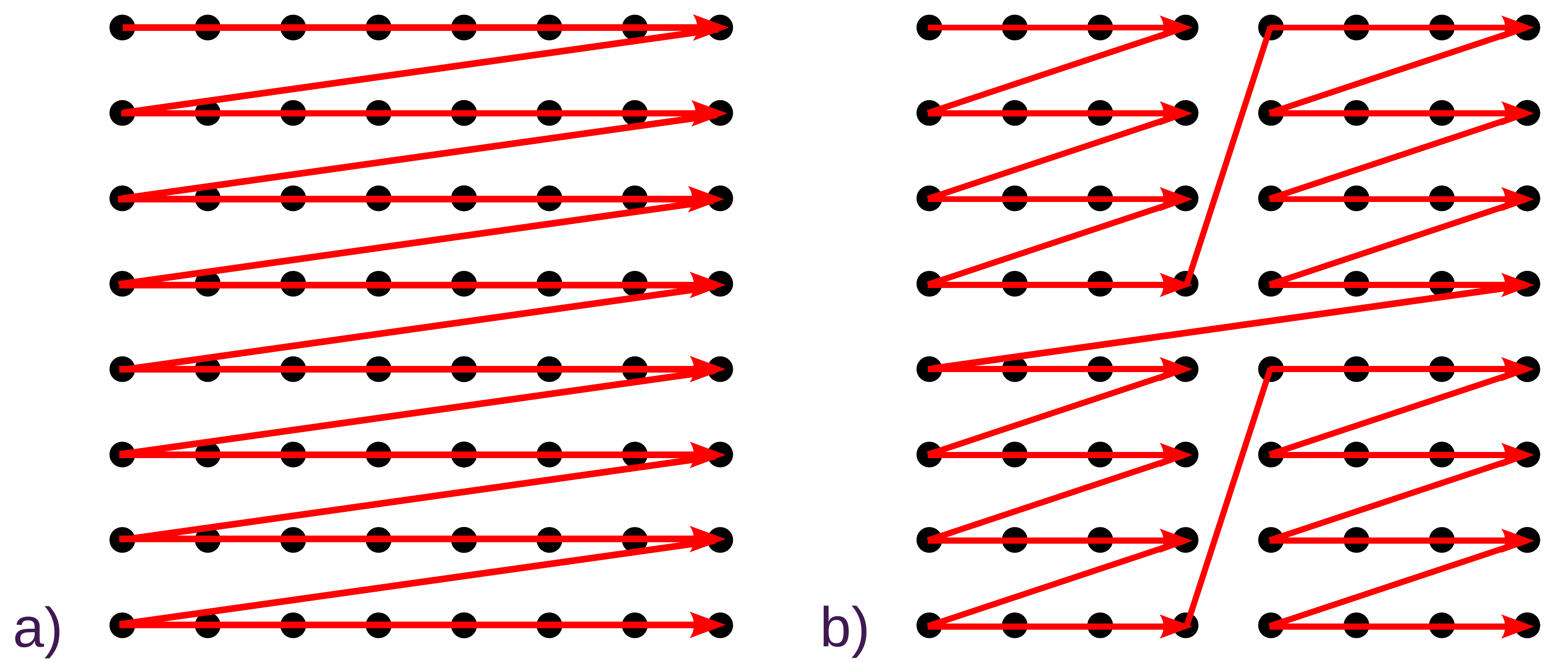}
  \caption{Examples of different grid orders in 2D: (a)~standard order
    (b)~grid ordered by small parallelepipedic blocks.}
  \label{fig:locality}
\end{figure}

\begin{figure}[ht]
  \centering
  \includegraphics*[width=\columnwidth]{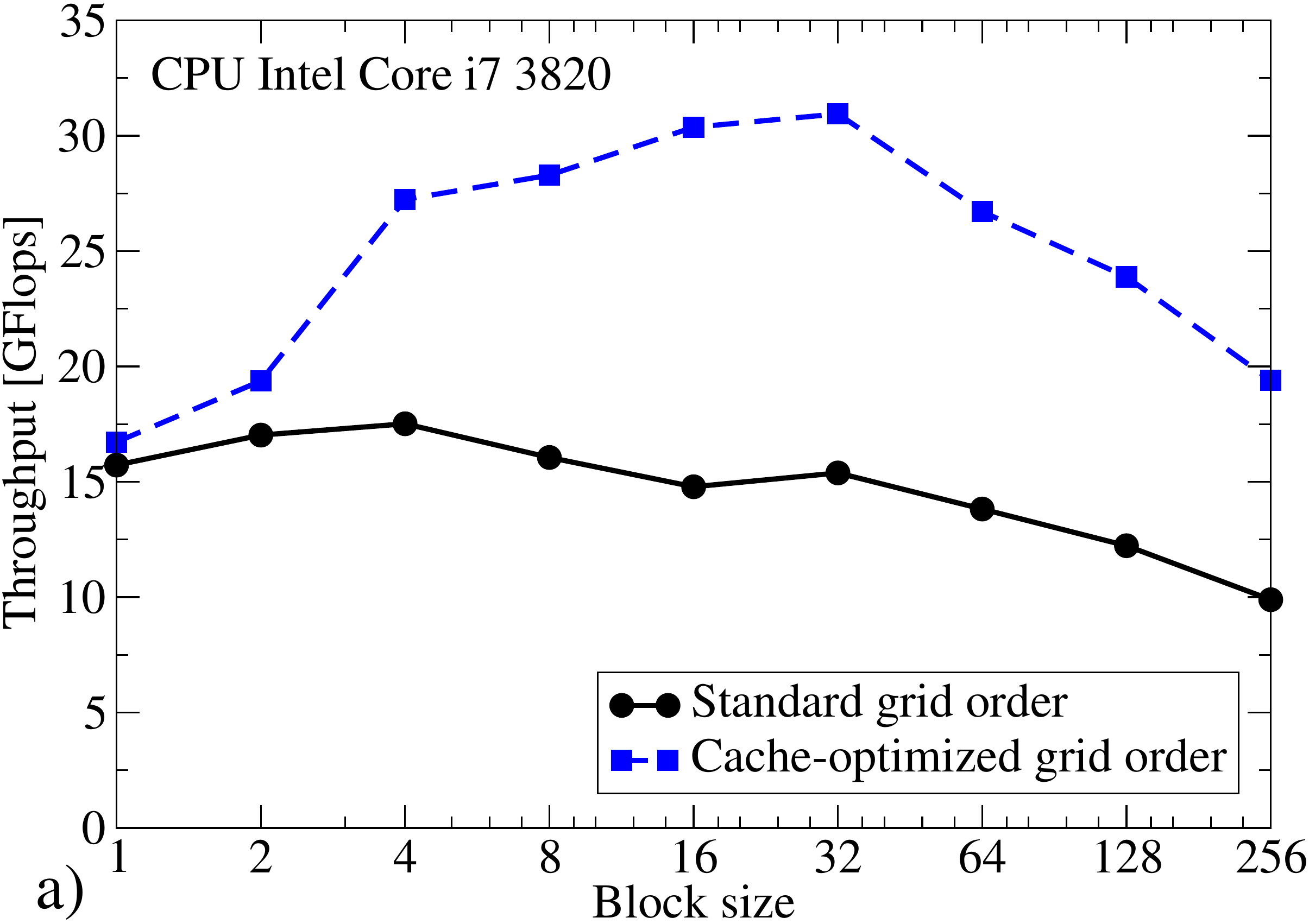}
  \includegraphics*[width=\columnwidth]{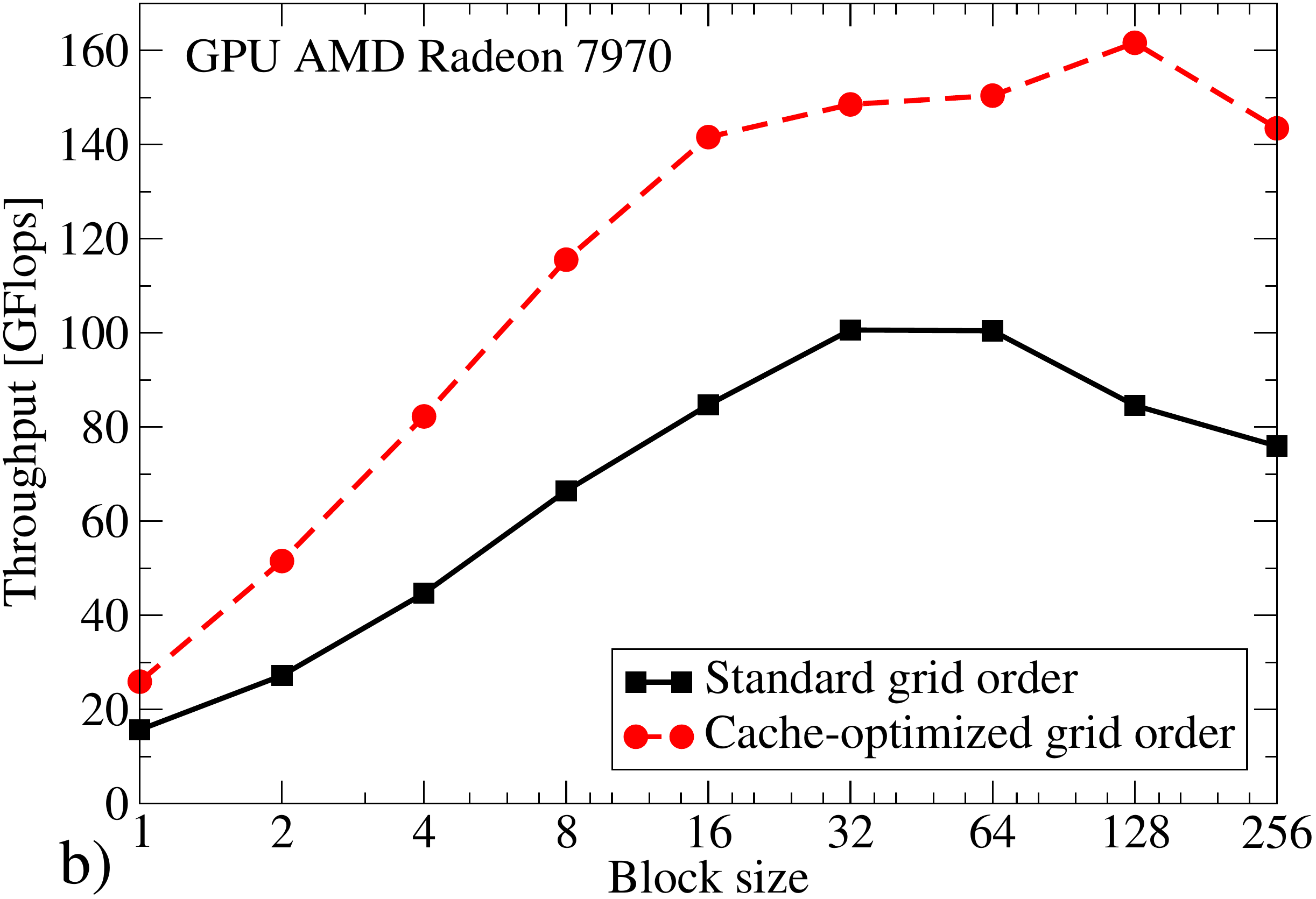}
  \caption{Effect of the optimization of the grid mapping for data
    locality in the numerical throughput of the Laplacian operator as
    a function of the size of the orbitals block. Spherical grid with
    ~500k points. a) computations with an Intel Core i7 3820 (8
    threads). b) computations with a AMD Radeon 7970.}
  \label{fig:cache}
\end{figure}

We devised an approach to improve cache utilization by controlling how
grid points are ordered in memory, \emph{i.e.}, how the
three-dimensional grid is mapped to a linear array. The standard
approach is to use a row-major or column-major order which leads to
some neighboring points being allocated in distant memory
locations. Our approach is to enumerate the grid points based on a
sequence of small parallelepipedic grids, as shown in the example of
Fig.~\ref{fig:locality}. This approach permits close spatial regions
to be stored closer in memory, improving memory locality for the
Laplacian operator. The effect of this optimization can be seen in
Fig.~\ref{fig:cache}, where we compare the throughput of the Laplacian
operator, as a function of the block-size, for the optimized grid
order with respect to the standard one. For the CPU with the standard
ordering of points, there is only a small gain performance from using
blocks of orbitals, while by optimizing the grid order, the
parallelism exposed by a larger block size allows a considerable
performance gain. For the GPU the effect of the optimization is less
dramatic but still significant.

\begin{figure}
  \centering
  \includegraphics[width=\columnwidth]{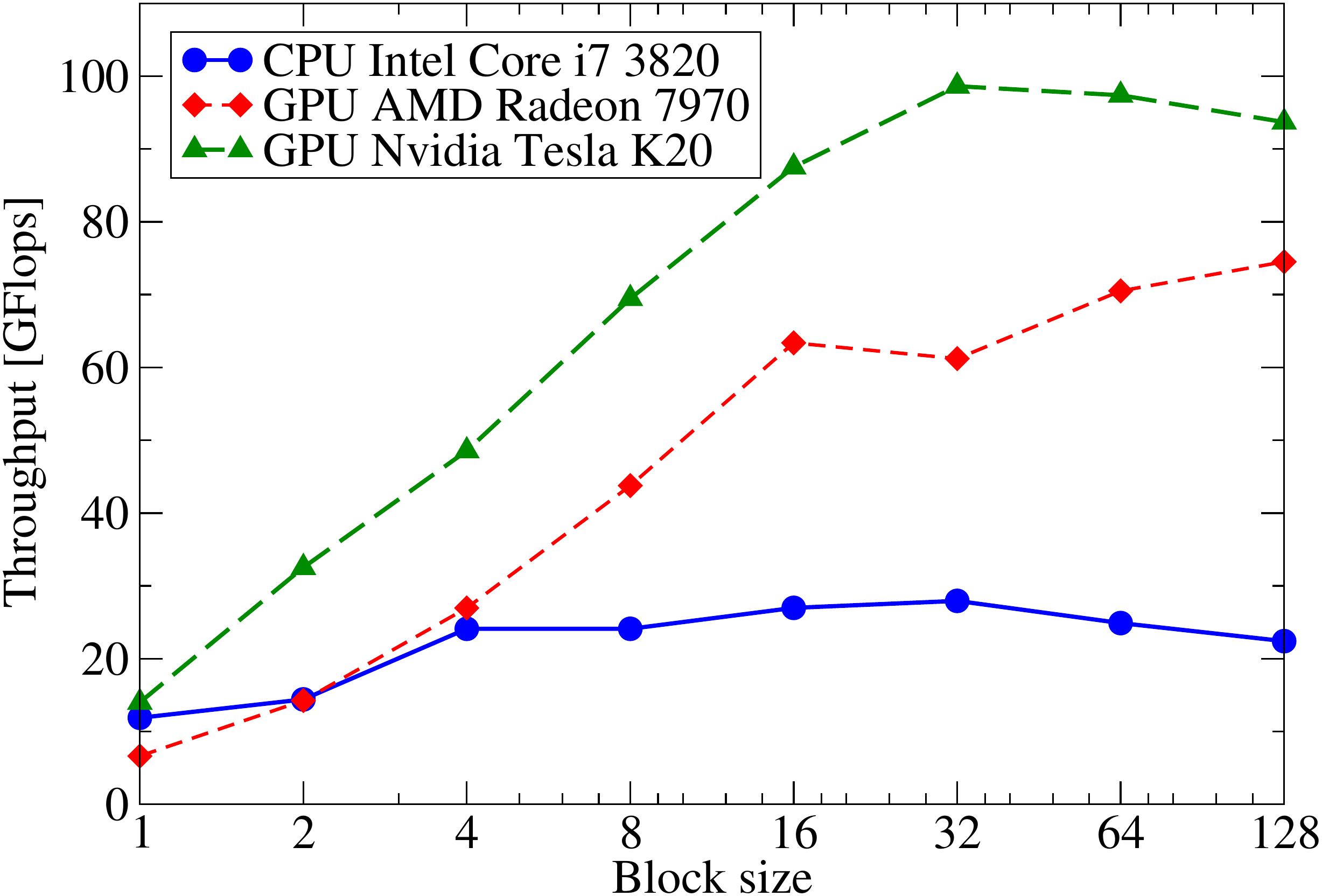}
  \caption{Numerical throughput of the calculation of the
    finite-difference fourth-order Laplacian as a function of the size
    of the block of orbitals (block-size) for different
    processors. Calculation for \(\beta\)-cyclodextrin with 256 orbitals
    and 260k grid points.}
  \label{fig:laplacian}
\end{figure}

An area for further improvement, is that the optimal size of the
parallelepipedic subgrids depends on the processor and the shape and
size of the grid, which change for each molecule. Since it is not
practical to optimize these parameters for each case, we use a fixed
set that does not always yield the best possible performance. This can
be seen in Fig.~\ref{fig:laplacian}, where we show a comparison of the
numerical throughput of the GPU and CPU implementations of the
Laplacian operator for a \(\beta\)-cyclodextrin molecule: the performance
obtained is not as high as in Fig.~\ref{fig:cache}. We plan to study
the applicability of more sophisticated space-filling
curves~\cite{Peano1980,*Sagan1994,*Gunther2006} to address this issue.

It is clear from Fig.~\ref{fig:laplacian} that for all processors, the
use of blocks of KS states represents a significant numerical
performance gain with respect to working with one state at a
time. This is particularly important for GPUs, where performance with
a single state is similar to the CPU, but it is more than five times
larger with blocks of size 32 or 64.

\subsection{Local potential}

The second term of the Hamiltonian is the local potential, that
includes contributions from the external potential, including the
local parts of the pseudo-potentials, the Hartree, exchange, and
correlation potentials. All these terms are summed into a single
potential, so we only need to multiply each orbitals by this potential
and store the result.

\begin{figure}[t]
  \centering
  \includegraphics[width=\columnwidth]{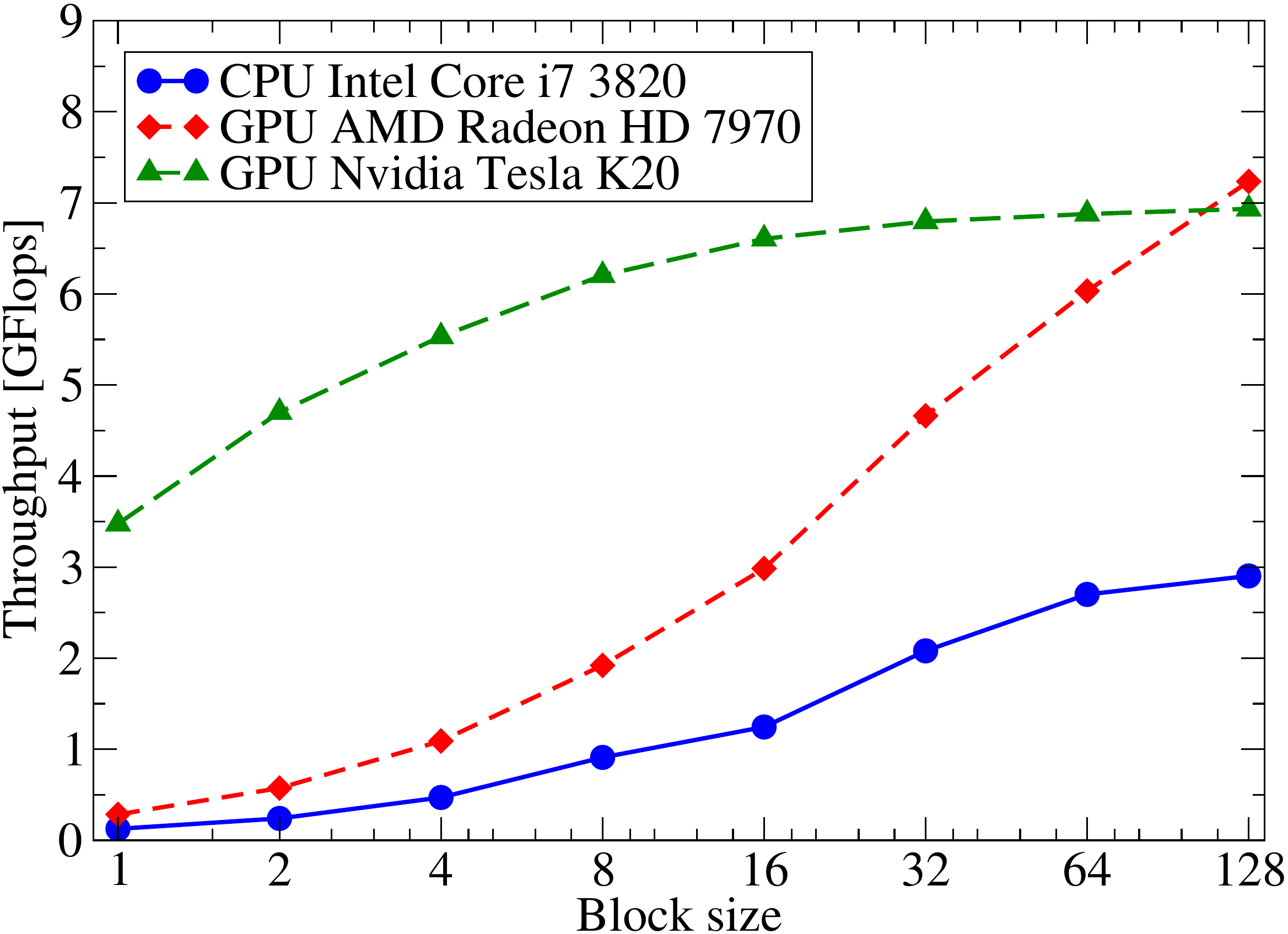}
  \caption{Numerical throughput of the application of the local
    potential as a function of size of the block of orbitals
    (block-size) for different processors. Calculation for
    \(\beta\)-cyclodextrin with 256 orbitals and 260k grid points.}
  \label{fig:vlocal}
\end{figure}

Since there are only two arithmetic operations per element, the
application of the local potential is heavily limited by memory
access. Using blocks of orbitals has two beneficial effects: the
larger number of simultaneous operations can hide the memory latency,
and the values of the potential are reused, reducing the number of
memory accesses. In Fig.~\ref{fig:vlocal}, we compare the numerical
performance of the application of the local potential for different
processors. As expected, the GPU has a considerable performance
advantage caused by the higher memory bandwidth. Still, the numerical
throughput is significantly below the values we obtain for other
parts of the calculation.

\subsection{Non-local potential}

The final term required for the application of the Hamiltonian is the
non-local potential that comes from the norm-conserving
pseudo-potentials~\cite{Kleinman1982,*Troullier1991}. The non-locality
comes from the fact that each angular momentum component of the
orbital sees a different potential. In practice, we calculate
\begin{equation}
  \label{eq:pseudo}
  V_\mathrm{nl}\varphi_k(\vec{r}) =
  \sum_A\sum_{lm}\gamma^A_{lm}(\vec{r} - \vec{R})
  \int_{r'<r_c}\!\!\!\!\!\!\!\!\!\mathrm{d}\vec{r}'\,\gamma^A_{lm}(\vec{r}'
  - \vec{R}_A)\varphi_k(\vec{r}')\,,
\end{equation}
where \(\gamma^A_{lm}\) corresponds to the pseudo-potential projectors
for atom \(A\), and \(l\) and \(m\) are the angular momentum
components that go from 0 to a certain \(l_\mathrm{max}\), usually 3,
and from \(-l\) to \(l\), respectively. The projector functions are
localized over a sphere, such that \(\gamma^A_{lm}(\vec{r}) = 0\) for
\(\left|\vec{r}\right| > r_\mathrm{c}\).

In our implementation, eq.~(\ref{eq:pseudo}) is calculated in two parts
that are parallelized differently on the GPU. The first part is to
calculate the integrals over \(\vec{r'}\) and store the results. 
This calculation is parallelized for a block of orbitals,
angular-momentum components and all atoms, with each GPU-thread
calculating an integral.

\begin{figure}[t]
  \centering
  \includegraphics[width=0.6\columnwidth]{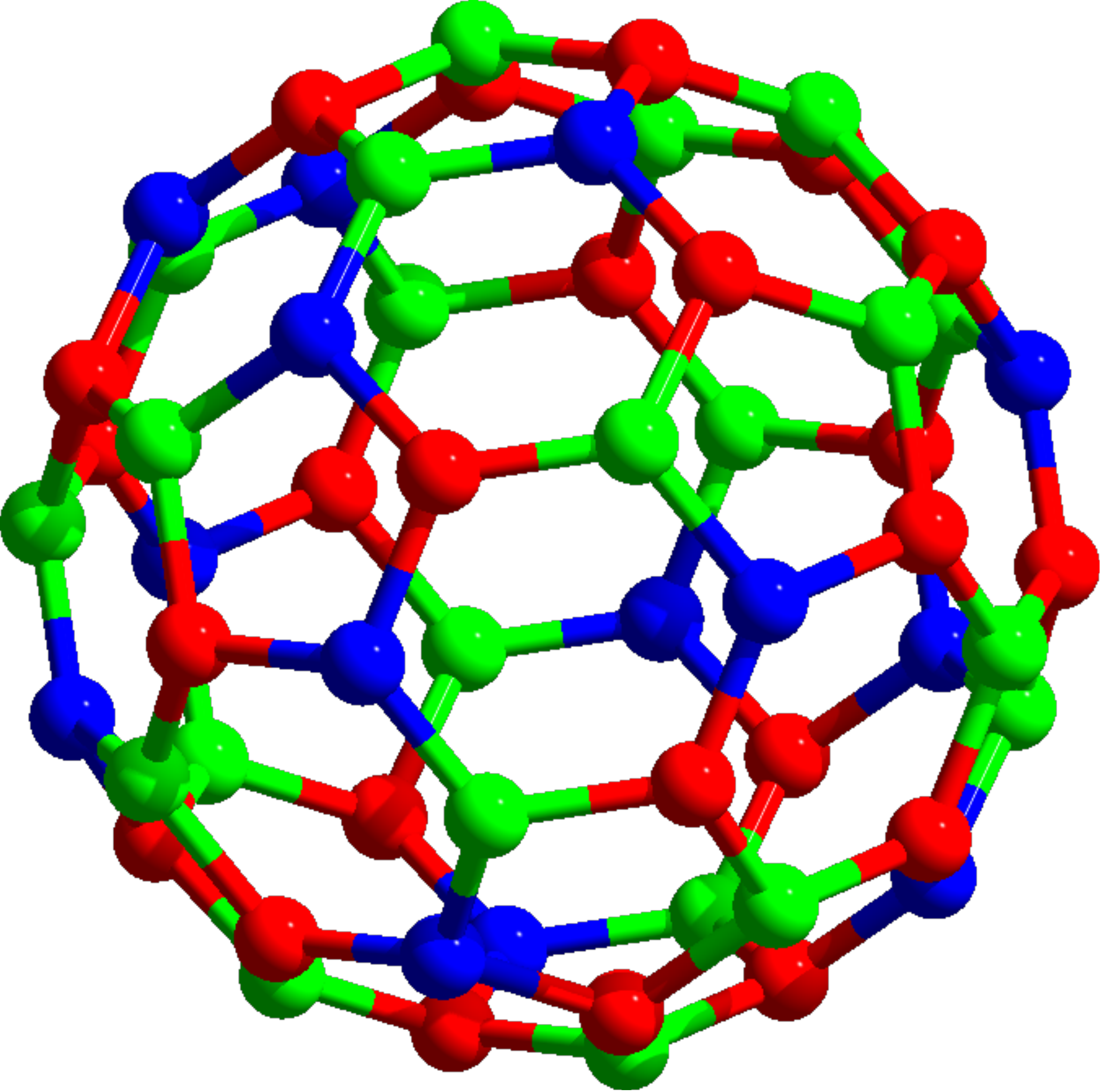}
  \caption{Division of the atoms of a C\s{60} molecule in groups
    (represented by different colors) whose pseudo-potential spheres
    do not overlap.}
  \label{fig:colors}
\end{figure}

The second part of the application of the non-local potential is to
multiply the stored integrals by the radial functions and sum over
angular-momentum components.
In this case, the calculation can be parallelized over orbitals, and,
if the pseudo-potential spheres associated to each atom do not
overlap, it can also be parallelized over the \(\vec{r}\)-index and
atoms. Usually the spheres do not overlap, but if they do, race
conditions would appear as several threads would try to update the
same point. In order to do the calculations in parallel, we divide the
atoms in groups whose spheres do not overlap. Then, we parallelize
over all atoms in each group. In Fig.~\ref{fig:colors} we show an
example of the division of atoms for the C\(_{60}\) molecule in
non-overlapping groups.

\begin{figure}[t]
  \centering
  \includegraphics[width=\columnwidth]{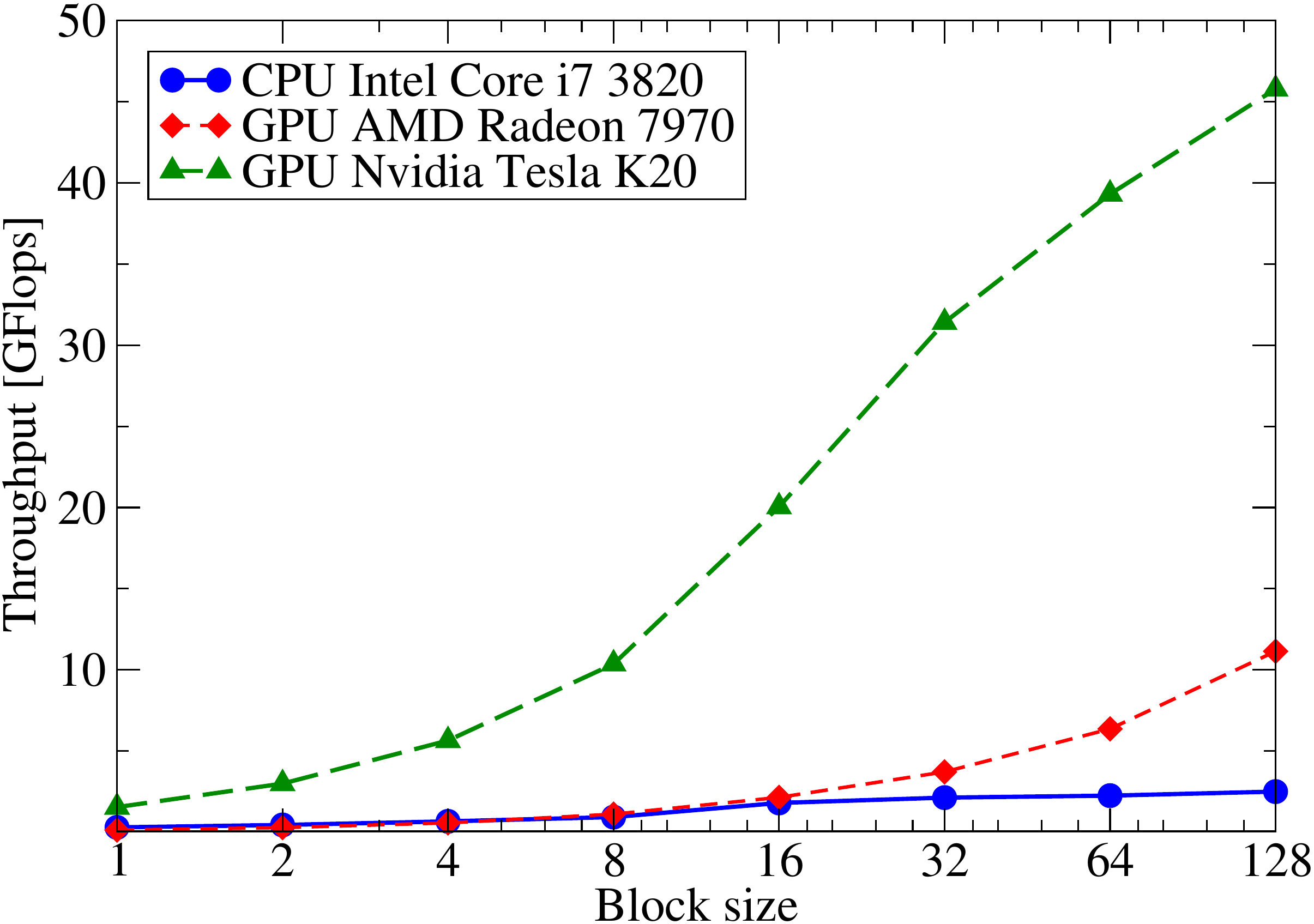}
  \caption{Numerical throughput of the application of the non-local
    potential as a function of the size of the block of orbitals
    (block-size). Calculation for \(\beta\)-cyclodextrin with 256 orbitals
    and 260k grid points.}
  \label{fig:nonlocal}
\end{figure}

In Fig.~\ref{fig:nonlocal}, we plot the throughput obtained by the
non-local potential implementation for a \(\beta\)-cyclodextrin
molecule. The Nvidia card shows a good performance, 46 GFlops, only
when large blocks of orbitals are used. The AMD card has a similar
behavior, but the performance is much lower, with a maximum of 11
GFlops. This is a clear example of how our approach is an effective
way of increasing the performance that can be obtained from the
GPU. As this is a complex routine, and our current implementation is
very basic, we suspect that a more sophisticated and optimized version
could significantly increase the numerical performance of this part of
the application of the KS Hamiltonian, in particular for the AMD GPU.

\section{Orthogonalization and subspace diagonalization}
\label{sec:ortho}

Given a set of orbitals, \(\left\{\varphi_k\right\}\), the
orthogonalization process generates a new set of orthogonal orbitals,
\(\left\{\bar\varphi_k\right\}\), as a linear combination of the
original ones. Our implementation of the orthogonalization procedure
is based on the Cholesky decomposition and other matrix linear algebra
operations~\cite{Kresse1996}. For CPUs \textsc{blas} and
\textsc{lapack} provide an efficient and portable set of routines to
perform these operations. For GPUs, we use the OpenCL \textsc{blas}
implementation provided by AMD as part of the Accelerated Parallel
Processing Math Libraries (APPML).

The first step of the orthogonalization is to calculate the overlap
between orbitals, 
\begin{equation}
  \label{eq:overlap}
  S_{jk} = \langle\varphi_j| \varphi_k\rangle \ .
\end{equation}
Our first approximation to this problem was to use the orbitals-block
approach to calculate the matrix \(S\), by dividing it into
sub-matrices, where each sub-matrix corresponds to the dot product
between all the elements of two blocks of orbitals; however, this
scheme is not efficient as it reduces the amount of data reuse in the
matrix multiplication~\cite{Wadleigh2000}.

We have found that a much more efficient approach is to first copy the
data to an array where all the coefficients corresponding to different
orbitals are contiguous in memory, then we can use \textsc{blas} to
calculate \(S\) as a \emph{rank k} operation. To avoid allocating a
full copy of all the orbitals, we perform the operation for a set of
points at a time. Effectively, we are switching from a
block-of-orbitals representations to a block-of-points approach.  Once
\(S\) is calculated, we need to factorize it into a \(U^\dagger U\)
form using a Cholesky decomposition~\cite{Benoit1924}. In our
implementation, this operation is done on the CPU using
\textsc{lapack}\footnote{The \textsc{magma} project~\cite{Agullo2009}
  implements some of the \textsc{lapack} calls on OpenCL, including
  the Cholesky decomposition \review{and dense matrix diagonalization}. We
  expect to support this library in the future.}. However, this is not
an issue in our current implementation, since the cost of the
decomposition is much smaller than other operations.

From the upper-triangular matrix \(U\), given by the Cholesky decomposition, we
can obtain the new set of orthogonal orbitals from the linear equation
\begin{equation}
  \label{eq:trsm}
  \sum_{k}U_{jk}\bar{\varphi}_k(\vec{r}) = {\varphi}_j(\vec{r})\ .
\end{equation}
Since \(U\) is triangular, the solution of the linear problem is a
simple operation that is done by \textsc{blas}. As this procedure
mixes all states we cannot use the blocks-of-orbitals approach,
instead we switch again to the blocks-of-points representation.

\begin{figure}
  \centering
  \includegraphics[width=\columnwidth]{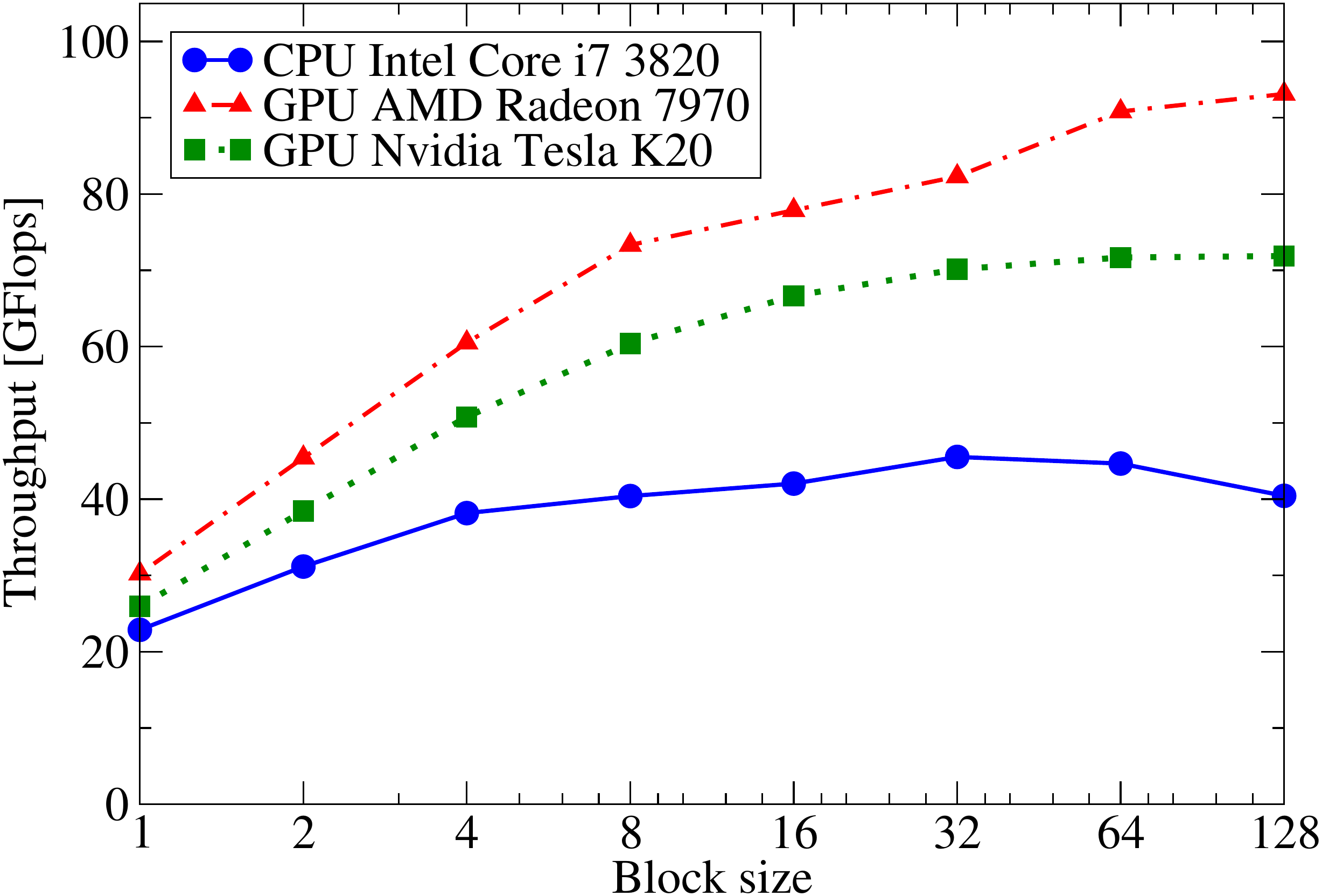}
  \caption{Numerical throughput of the orthogonalization procedure as
    a function of the size of the block of orbitals (block-size) for
    different processors. Calculation for
    \(\beta\)-cyclodextrin with 256 orbitals and 260k grid points.}
  \label{fig:ortho}
\end{figure}

In Fig.~\ref{fig:ortho} we show the performance obtained for our
implementation of the orthogonalization procedure. The GPU speed-up is
not very large with respect to the CPU. As this operation is based on
linear algebra operations, we attribute the poor speed-up to
difference in the linear algebra libraries. While for CPUs
\textsc{blas} implementations are quite mature, the implementation of
linear algebra operations on a GPU is still a field of active study,
in particular for the solution of triangular systems~\cite{Ries2012},
like eq.~(\ref{eq:trsm}).

The procedure for subspace diagonalization is very similar in form to
the orthogonalization. It is used by diagonalization algorithms for
sparse matrices to resolve between eigenvectors that have close
eigenvalues. The first step in subspace diagonalization is to generate
the representation of the Hamiltonian in the subspace of the
approximated orbitals, \(\left\{\varphi_k\right\}\),
\begin{equation}
  \label{eq:hsubspace}
  h_{jk} = \langle\varphi_j| H |\varphi_k\rangle \ .
\end{equation}
As in the case of the matrix \(S\) for the orthogonalization, we
perform this operation by blocks of points. This time we need to apply
the Hamiltonian to the orbitals first, and then calculate the dot
products as a matrix multiplication.

Once the subspace Hamiltonian is calculated, it is diagonalized to
obtain the matrix of its eigenvectors, \(\xi_{jk}\). As in the case of
the Cholesky decomposition, this dense-matrix diagonalization is done
by the CPU. This is not a performance issue for the systems studied in
this article, but for larger systems, the dense eigensolver, that
scales as \(O(n^3)\), could consume a considerable part of the
computation time.

Once the subspace Hamiltonian is diagonalized, the new set of
orbitals, \(\left\{\bar\varphi_k\right\}\), is generated by rotating
the old set by the eigenvector matrix,
\begin{equation}
  \label{eq:rotation}
  \bar{\varphi}_k(\vec{r}) = \sum_j\xi_{jk}\,{\varphi}_j(\vec{r})\ .
\end{equation}
Since this rotation mixes all orbitals, we follow a similar procedure
as we do in eq.~(\ref{eq:trsm}) for the orthogonalization. The only
difference is that in this case we directly multiply by the matrix
instead of its inverse.

\begin{figure}
  \centering
  \includegraphics[width=\columnwidth]{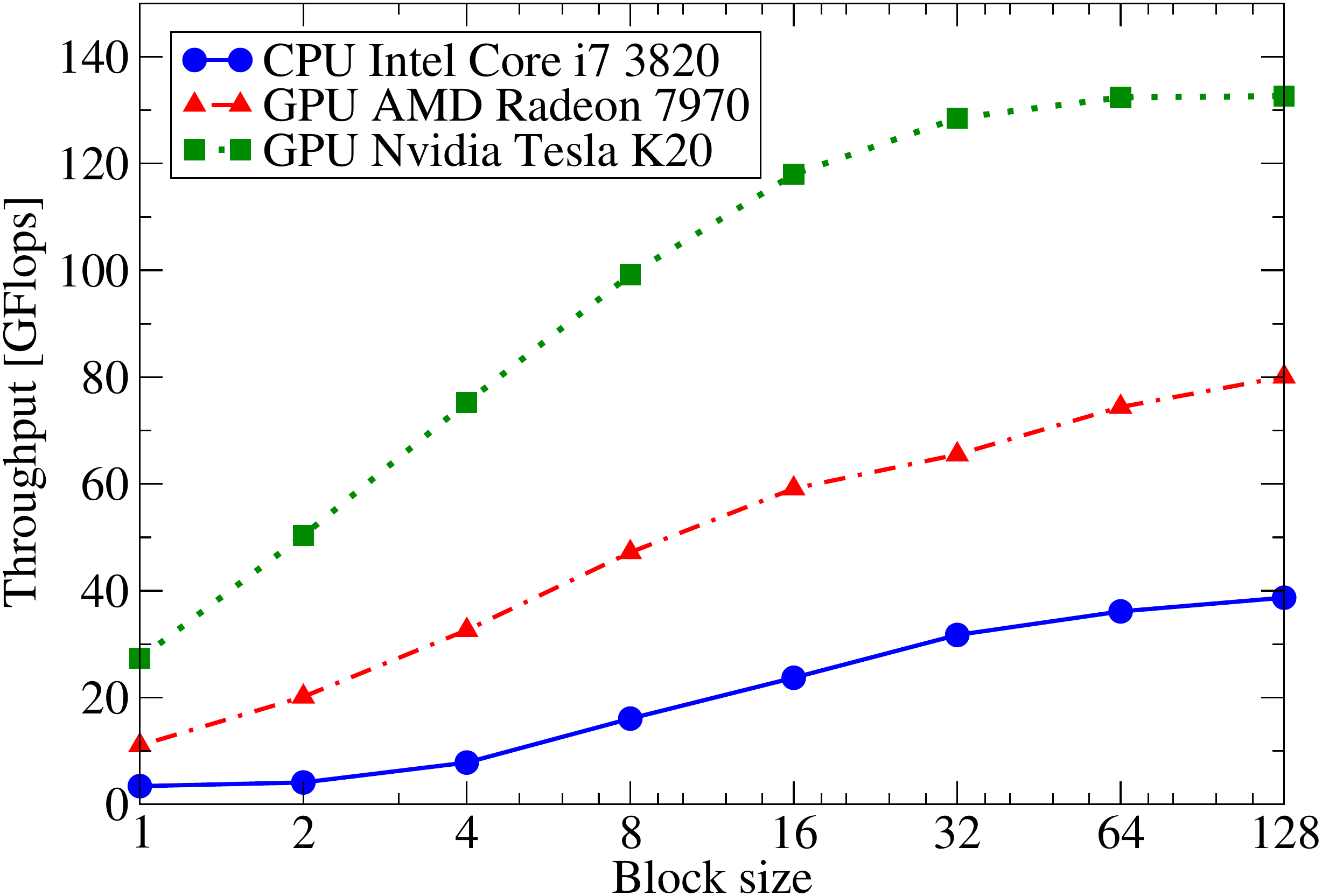}
  \caption{Numerical throughput of the subspace-diagonalization
    procedure as a function of the size of the block of orbitals
    (block-size) for different processors. Calculation for
    \(\beta\)-cyclodextrin with 256 orbitals and 260k grid points.}
  \label{fig:subspace}
\end{figure}

Fig.~\ref{fig:subspace} shows the performance obtained for the
subspace diagonalization. In this case the GPU speed-up is larger than
for the orthogonalization case, probably because this routine is based
on our implementation of the KS Hamiltonian, and on matrix-matrix
multiplications, that in general are simpler to optimize and parallelize
than other linear algebra operations.

\section{The Hartree potential}
\label{sec:poisson}

Other operation that we execute on the GPU is the calculation of the
Hartree potential by solving the Poisson problem,
eq.~(\ref{eq:poisson}). This equation also appears in other contexts
in electronic structure simulations, for example, in the calculation
of approximations to the exchange
term~\cite{Perdew1981,*Umezawa2006,*Andrade2011}, in the calculation
of integrals that appear in Hartree--Fock or Casida
theories~\cite{Shang2010}, or to impose electrostatic boundary
conditions~\cite{Tan1990,*Luscombe1992,Klamt1993,*Tomasi1994,Olivares-Amaya2011,*Watson2012}.

The Poisson equation can be solved by different methods in linear or
quasi-linear time~\cite{Greengard1997,*Kutteh1995,Briggs1987,*Beck1997,Cerioni2012,GarciaRisueno2013}. In our GPU implementation
we use an approach based on fast Fourier transforms (FFTs), as it is
quite efficient and simple to implement. By using FFTs, in principle
we are imposing periodic boundary conditions to the electrostatic
potential. We can, however, find the free-boundary solution by
enlarging the FFT grid and using a modified interaction
kernel~\cite{Rozzi2006}.

The solution process involves several steps. The first one is
to copy the density from the arbitrarily-shaped grid to a cubic grid,
where we perform the forward FFT. The result is the density in Fourier
space, that is multiplied by the Coulomb-interaction kernel. After an
inverse FFT, we obtain the Hartree potential, that is copied back to
the arbitrarily-shaped grid. Since we only need to solve a single
Poisson equation, independently of the size of the system, we cannot
use the block approach in this case. The essential component of this
solver is an FFT implementation, for GPUs, we use the clAMDFft library
provided by AMD. For CPUs we use the multi-threaded FFTW
library~\cite{Frigo2005}.

\begin{figure}
  \centering
  \includegraphics[width=\columnwidth]{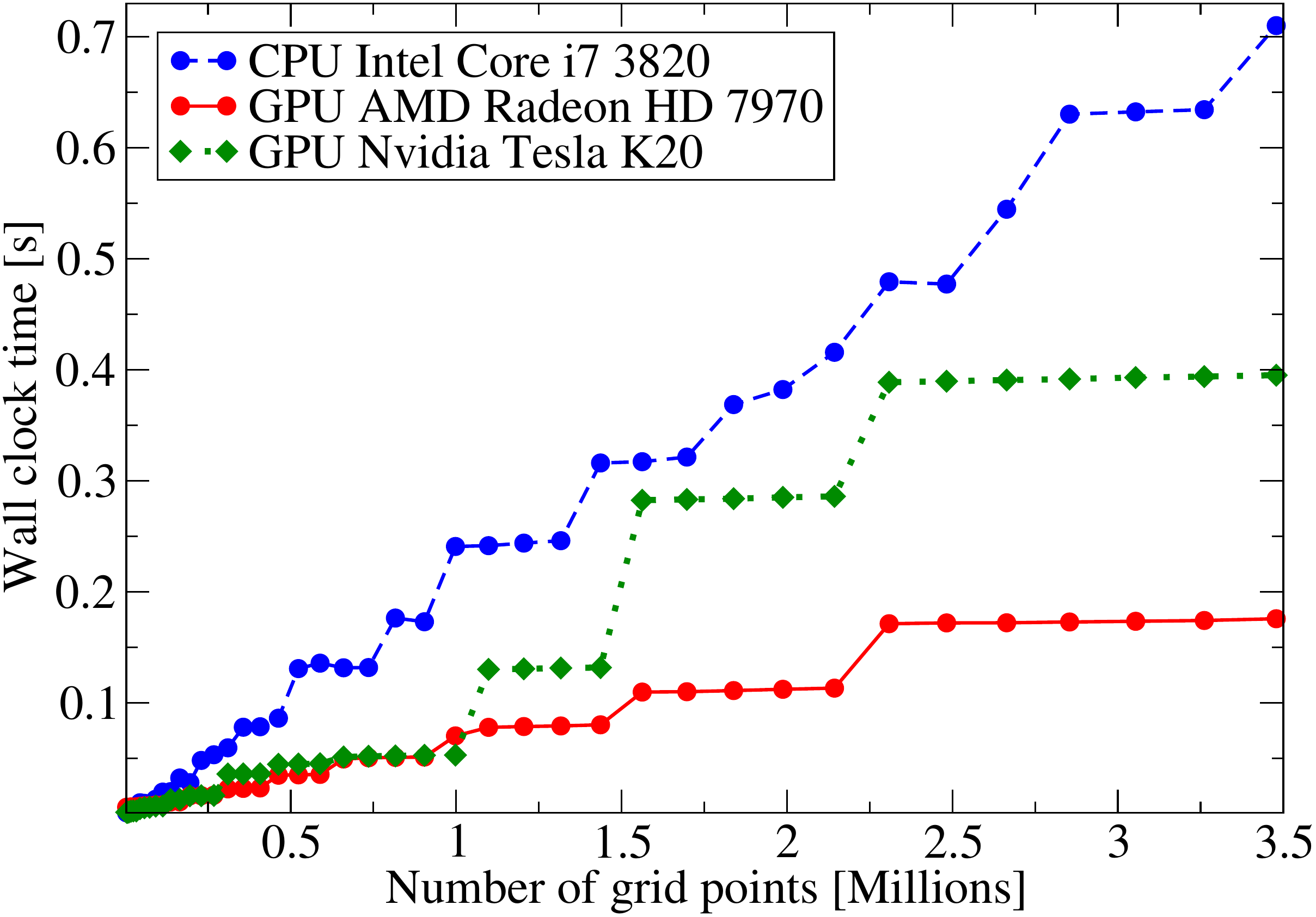}
  \caption{Comparison of the computational time required for solving
    the Poisson equation using FFTs as a function of the number of
    grid points. The data is originally on main memory, so the time
    required to copy the input data to the GPU and copy back the
    result is included. The number of points corresponds to the grid
    used by \textsc{octopus}, the FFT grid has a larger number of points.}
  \label{fig:poisson}
\end{figure}

In Fig.~\ref{fig:poisson}, we show the performance of our GPU based
Poisson solver for different system sizes. For the AMD card, the GPU
version outperforms the CPU version, in some cases by a factor of
7. For the Nvidia GPU the speed-up is smaller, probably because the
library has not been explicitly optimized for this GPU. The step
structure seen on the plot is caused by the fact that FFTs cannot be
performed efficiently over grids of any size: the grid dimension in
each direction must be a product of certain values, or radices, that
are determined by the implementation. If a grid dimension is not
valid, the size of the grid is increased. Since the CPU implementation
is more mature and supports more radices, the steps are smaller than
the GPU implementation that only supports radices 2 and
3~\footnote{The clAMDFft library also supports radix-5, but we could
  not use it due to execution and performance issues.}. So, it is
reasonable to expect that as the GPU-accelerated FFT implementations
improve, the numerical performance of the calculation of the Hartree
potential will increase.

\section{Other operations}

In the previous sections we have described the main operations that we
have implemented on the GPU. There are several simpler operations that
also need to be performed on the GPU. These operations include basic
operations between orbitals, like copies, linear combinations, and dot
products.  All of them are implemented on the GPU using the
block-of-orbitals approach to improve performance. In fact, we have
found that it is necessary to pay attention to the parallelization of
most of the operations performed on the GPU, as a single routine that
is not properly parallelized can spoil the numerical performance of
the entire code.

In our current implementation, there are two procedures that are still
done by the CPU, as they would require a considerable effort to
implement on the GPU, but have a minor impact in numerical
performance. The first one is the evaluation of the XC potential. This
is a local operation that is \review{straightforward} to parallelize
and should perform well on the GPU. The problem is that there is large
number of XC approximations, each one involving complex
formulas~\cite{Marques2012} that would need to be implemented on the
GPU. The second procedure that is executed on the CPU is the
initialization of the molecular orbitals by a linear combination of
the atomic orbitals obtained from the pseudo-potentials. The reason is
that we use a spline interpolation to transfer the orbitals to the
grid, which depends on the GSL library~\cite{gsl} that is not
available on the GPU.

\section{Accuracy}

The strategy presented in this article does not imply a reduction in
the precision of the calculations with respect to the original
real-space DFT implementation. There are, however, some factors that
could produce some numerical differences in the results.

In a sparse eigensolver, usually the eigenvectors are only converged
until their error goes below a certain threshold, to avoid wasting
computational time in over-converging some eigenvectors. In our
implementation, iterations are only stopped when a whole block of
eigenvectors is below the threshold. This makes the code simpler and
avoids thread divergence, but introduces a dependency of the results
on the block-size.

Another source of differences in the results is the calculation of the
Hartree potential. Since the number of prime factors supported by the
GPU FFT library is smaller than the CPU implementation, the size of
the FFT grid can be larger for the GPU. However, as in both cases the
grid is large enough to eliminate periodicity effects, the change in
the results due to this difference is minimum.

Finally, there might be some differences in the numerical
operations. While we use double precision for all operations and both
GPUs used for the tests are IEEE-754 compliant, there might be
differences in the finite precision arithmetic from fused multiply
addition (FMA) operations, \review{that are not available in the
  tested CPU}, and due to different ordering of operations.

In our tests with different molecules we observe that the difference
in the total energy between CPU and GPU calculations is on average 0.1
\review{millihartree} with a maximum of 0.5
\review{millihartree}. \review{This is difference is caused mainly by
  the different size of FFT grids used by the CPU and GPU
  implementations of the Poisson solver}. The difference between the
energy computed with the Nvidia GPU with respect to the AMD GPU is on
average 0.008 \review{millihartree} with a maximum of 0.08
\review{millihartree}. The variation of the total energy with the
block-size is well below this values.

\section{Numerical performance}




In this section we evaluate the numerical performance of our
implementation and how it depends on the size of the blocks of
orbitals or the size of the molecular systems. For this analysis we
use several parameters: the throughput, the total calculation time for
a single-point energy calculation, the speed-up with respect to the
CPU implementation, and the comparison with a second GPU
implementation.






\subsection{Block-size}

\begin{figure}
  \centering
  \includegraphics[width=\columnwidth]{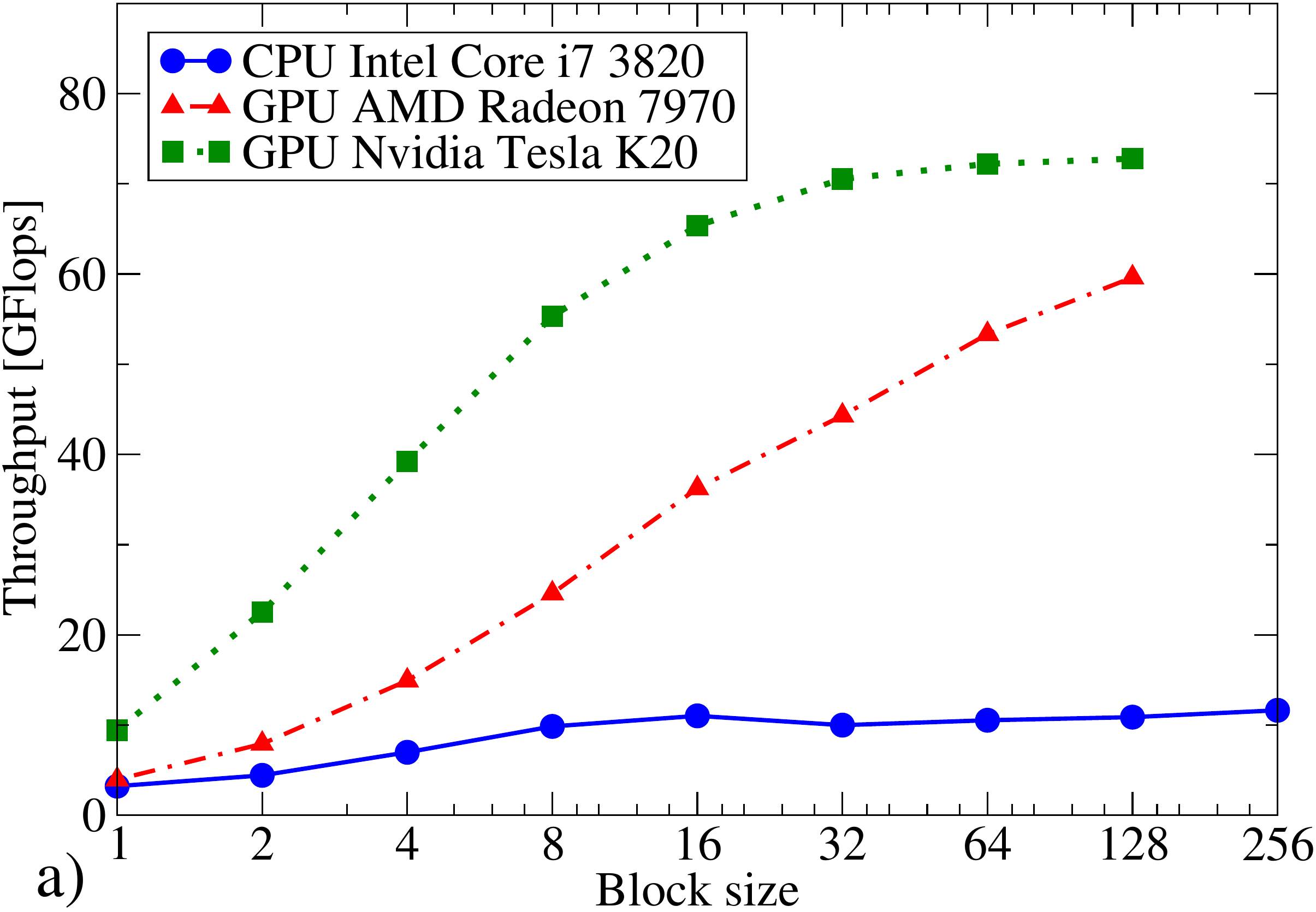}
  \includegraphics[width=\columnwidth]{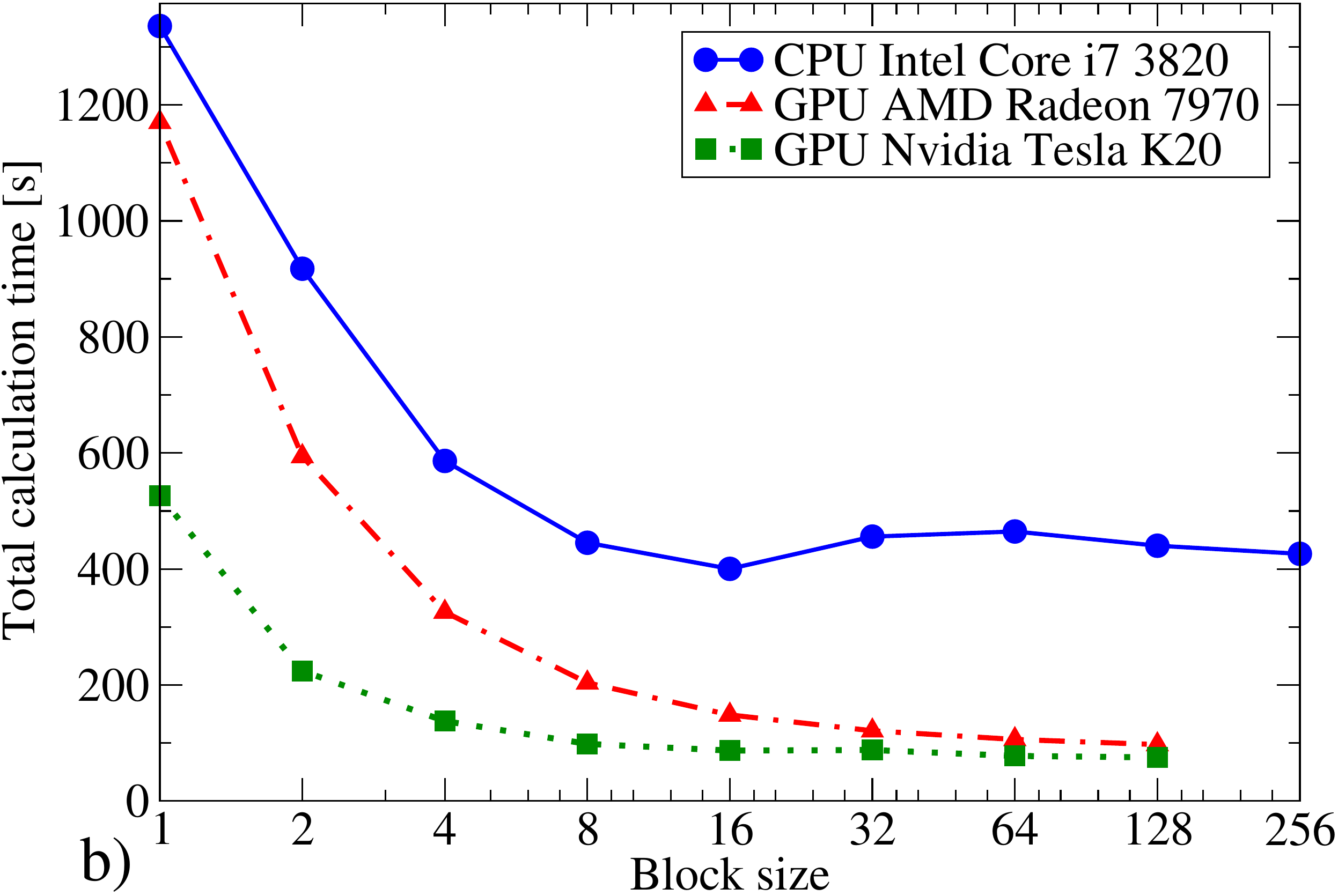}
  \caption{Performance of our CPU and GPU implementations as a
    function of the size of the block of orbitals (block-size). a)
    Numerical throughput of the self-consistency cycle. b) Total
    execution time for a single-point energy calculation. Simulation
    for \(\beta\)-cyclodextrin with 256 orbitals and 260k grid points.}
  \label{fig:blocksize}
\end{figure}

We start our performance analysis by studying how the block-size
influences execution performance. In Fig.~\ref{fig:blocksize} we plot,
for the \(\beta\)-cyclodextrin molecule, both the numerical throughput
obtained for the SCF loop and the total execution time as a function
of the size of the blocks of orbitals. For the CPU the optimal
block-size is 16, with a second local optimum for block-size 256. For
GPUs, increasing the block-size always improve performance up to size
128, that is the limit imposed by the GPU memory. This shows how the
block approach produces a significant improvement with respect to
working with a single orbital at a time (the block-size 1 case).


\subsection{Molecule size}
\label{sec:performance}

We now focus our attention on how our GPU implementation performs for
molecules of different sizes. For this test we have selected a set of
40 molecules, listed in table~\ref{tab:molecules}. In this respect, we
would like to assert that we did \textit{not} select the set of
molecules based on any performance-related criterion, we just aimed to
have a set of molecules composed mainly of first and second row
elements with different numbers of valence electrons and that could
fit in the memory of our GPUs.

\begin{figure}
  \centering
  \includegraphics[width=\columnwidth]{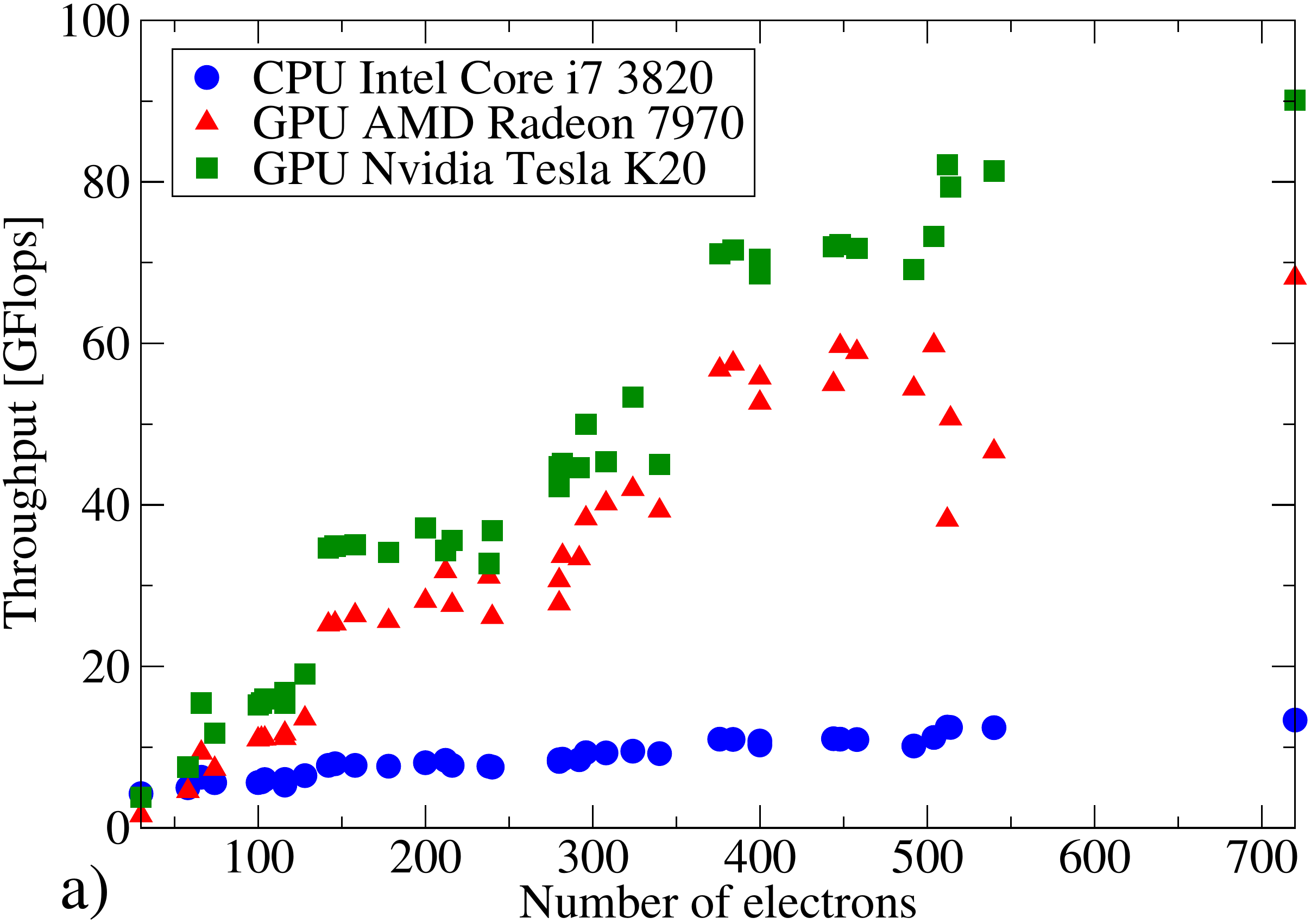}
  \includegraphics[width=\columnwidth]{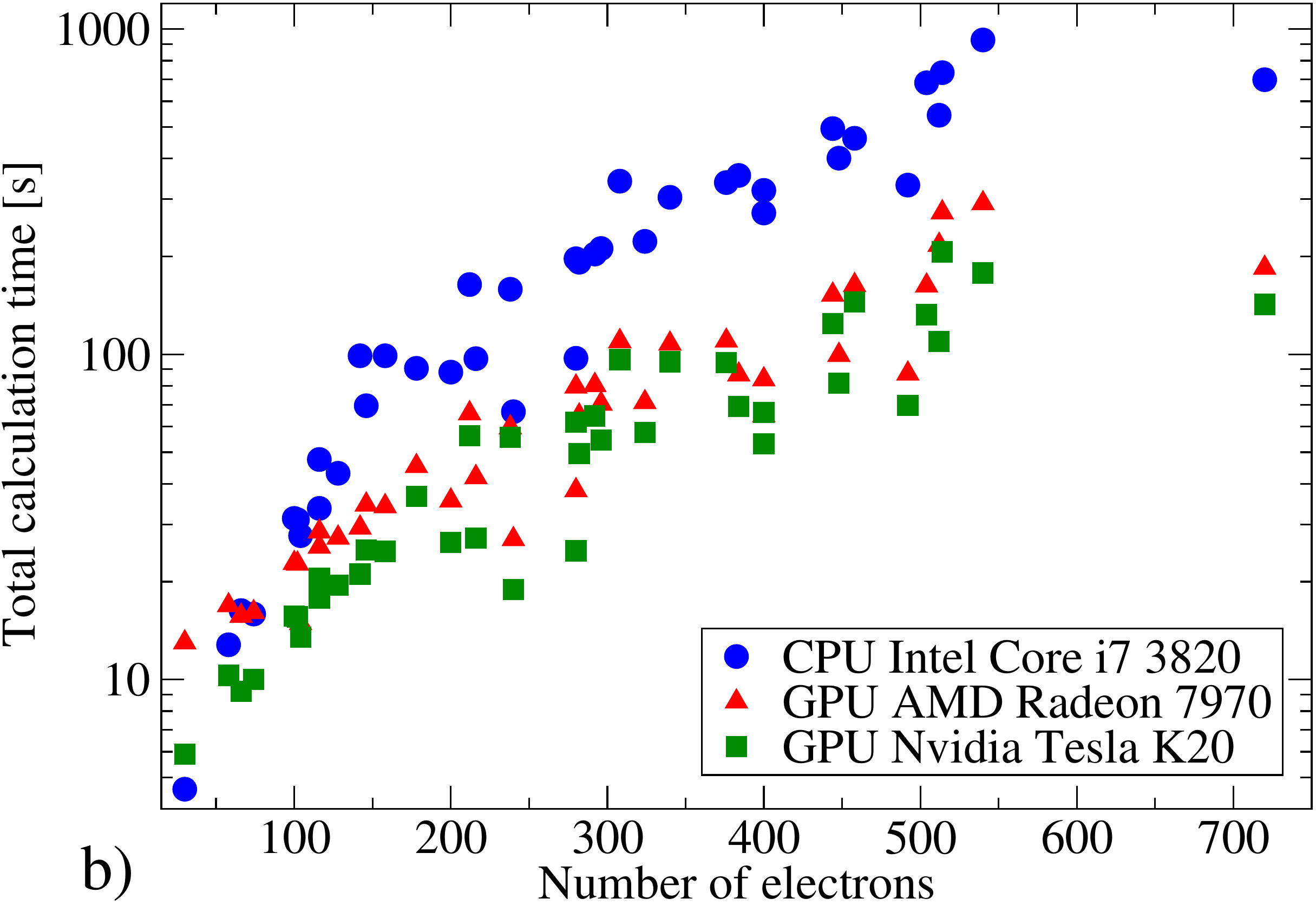}
  \caption{Performance of our CPU and GPU implementations for a set of
    40 molecules of different sizes. a) Numerical
    throughput of the self-consistency cycle. b) Total
    execution time for a single-point energy calculation. The list of
    molecules and the calculation times are given in table
    \ref{tab:molecules}.}
  \label{fig:molecules}
\end{figure}

\begin{table*}[p]
  \footnotesize
  \centering
  \caption{List of systems used for the performance studies done in
    this article. For each molecule we include the number of valence
    electrons, the number of grid points used in the simulation, the computational times for \textsc{octopus} with three different
    processors: CPU Intel Core i7 3820 \review{(CPU)}, GPU AMD Radeon HD 7970
    \review{(AMD)} and GPU Nvidia Tesla K20 \review{(Nvidia)}, and calculation time for \textsc{terachem} with
    the Nvidia Tesla K20 GPU \review{(Terachem)}. The geometry for each molecule can
    be found as supplementary information. \(^a\)These molecules were obtained
    from the Harvard Clean Energy Project (CEP)~\cite{Hachmann2011}.\label{tab:molecules}}
  \begin{tabular*}{\textwidth}{@{\extracolsep{\fill}} llrrrrrr}
    \multicolumn{2}{c}{System} & \multicolumn{2}{c}{Calculation size} &
    \multicolumn{4}{c}{Single-point calculation time \review{[s]}}\\
    Stoichiometry       	&	  Description             	&
    Electrons	&
    \review{Points [1/1000]} &	CPU	&	\review{AMD}	&	\review{Nvidia} & \review{Terachem}\\
    \hline	
    C\s{6}H\s{6}	&	benzene	&	30	&	37.3	&	4.6	&	13.0	&	5.9	&	2.0	\\
    C\s{10}H\s{18}	&	cis-decalin	&	58	&	62.5	&	12.8	&	16.9	&	10.3	&	6.8	\\
    C\s{14}H\s{10}	&	anthracene	&	66	&	63.0	&	16.3	&	15.7	&	9.2	&	7.6	\\
    C\s8H\s{10}N\s4O\s2	&	caffeine	&	74	&	63.1	&	15.9	&	16.1	&	10.0	&	8.7	\\
    C\s{16}H\s{24}O\s2	&	palmitoyl	&	100	&	93.2	&	31.3	&	22.8	&	15.7	&	15.1	\\
    C\s{18}H\s{24}	&	cis-retinal	&	102	&	96.5	&	31.0	&	22.7	&	15.6	&	17.4	\\
    (H\s2O)\s{13}	&	water cluster	&	104	&	83.7	&	27.7	&	20.9	&	13.5	&	8.2	\\
    C\s{20}H\s{24}O\s2	&	ethinyl estradiol	&	116	&	99.1	&	33.6	&	25.6	&	17.8	&	24.1	\\
    C\s{18}H\s{32}O\s2	&	linoleic acid	&	116	&	122.7	&	47.5	&	28.5	&	20.5	&	14.4	\\
    C\s{22}H\s{28}O\s2 	&	etonogestrel	&	128	&	107.3	&	43.1	&	27.3	&	19.5	&	30.0	\\
    C\s{26}H\s{16}O\s3S	&	molecule from CEP\(^a\)	&	142	&	110.0	&	62.7	&	29.3	&	21.1	&	24.6	\\
    C\s{29}H\s{20}N\s2	&	molecule from CEP\(^a\)	&	146	&	119.9	&	69.5	&	34.5	&	25.0	&	30.4	\\
    C\s{34}H\s{22}	&	diphenylpentacene	&	158	&	131.2	&	99.0	&	34.1	&	24.8	&	33.8	\\
    C\s{22}H\s{30}N\s6O\s4S 	& sildenafil citrate	&	178	&	137.8	&	90.6	&	45.3	&	36.6	&	41.3	\\
    CH\s4(H\s2O)\s{24}	&	methane + water	&	200	&	132.6	&	88.2	&	35.5	&	26.4	&	27.8	\\
    C\s{40}H\s{52}	&	carotene	&	212	&	206.0	&	163.9	&	65.8	&	56.2	&	42.1	\\
    C\s{48}H\s{24}	&kekulene	&	216	&	147.6	&	97.0	&	41.9	&	27.2	&	49.2	\\
    C\s{44}H\s{54}Si\s2	&	TIPS-pentacene	&	238	&	182.8	&	158.6	&	59.6	&	55.5	&	68.6	\\
    C\s{60}	&	fullerene	&	240	&	102.4	&	66.6	&	27.0	&	18.9	&	76.6	\\
    C\s{70}	&	fullerene	&	280	&	113.1	&	97.2	&	38.2	&	24.9	&	128.3	\\
    C\s{51}H\s{33}N\s5O\s3	&	porphyrin	&	280	&	209.8	&	196.7	&	79.5	&	61.8	&	105.4	\\
    C\s{58}H\s{32}S\s3	&	molecule from CEP\(^a\)	&	282	&	214.9	&	192.0	&	64.8	&	49.5	&	82.0	\\
    C\s{41}H\s{40}N\s8O\s8	&	carbazole complex	&	292	&	192.3	&	203.5	&	80.4	&	64.7	&	126.2	\\
    C\s{60}H\s{32}S\s4	&	DAT-thiophane dimer	&	296	&	213.7	&	211.6	&	70.4	&	54.5	&	78.9	\\
    C\s{42}H\s{83}NO\s8P	&	phosphatidylcholine	&	308	&	283.9	&	340.8	&	109.7	&	96.4	&	95.7	\\
    C\s{45}H\s{51}NO\s{15}	&	taxol	&	324	&	219.0	&	222.6	&	71.2	&	57.6	&	141.1	\\
    C\s{50}H\s{238}MgN\s4O\s5	&	chlorophyll	&	340	&	269.8	&	303.9	&	107.8	&	94.9	&	174.5	\\
    C\s{58}H\s{48}N\s8O\s{12}	&	methotrexate complex	&	376	&	238.1	&	337.4	&	110.1	&	94.3	&	135.4	\\
    C\s{36}H\s{60}O\s{30}	&	\(\alpha\)-cyclodextrin	&	384	&	222.9	&	354.9	&	86.6	&	69.2	&	89.6	\\
    C\s{100}	&	fullerene	&	400	&	160.8	&	272.2	&	64.9	&	53.1	&	194.2	\\
    C\s{60}(H\s2O)\s{20}	&	fullerene + water &	400	&	200.5	&	319.2	&	83.7	&	66.3	&	225.2	\\
    C\s{54}H\s{90}N\s6O\s{18}	&	valinomycin	&	444	&	293.6	&	494.3	&	152.5	&	124.5	&	185.6	\\
    C\s{42}H\s{70}O\s{35}	&	\(\beta\)-cyclodextrin	&	448	&	259.5	&	401.0	&	99.5	&	81.6	&	100.6	\\
    C\s{62}H\s{63}N\s{15}O\s{12}	&	methotrexate complex	&	458	&	265.4	&	461.6	&	163.1	&	145.0	&	331.1	\\
    C\s{122}H\s4	&	fullerene dimer	&	492	&	198.3	&	331.4	&	87.2	&	69.6	&	344.0	\\
    C\s{114}H\s{48}	&	graphite cluster	&	504	&	277.8	&	684.1	&	162.6	&	132.5	&	672.4	\\
    C\s{48}H\s{80}O\s{40}	&	\(\gamma\)-cyclodextrin	&	512	&	290.9	&	543.3	&	216.8	&	109.5	&	131.0	\\
    C\s{68}H\s{76}N\s{13}O\s{16}P	&	cAMP  complex	&	514	&	327.4	&	734.6	&	273.0	&	206.7	&	334.1	\\
    C\s{68}H\s{318}Na\s2O\s{20}P\s2	&	phospholipid	&	540	&	404.7	&	925.8	&	291.3	&	177.8	&	808.8	\\
    C\s{180}	&	fullerene	&	720	&	267.8	&	699.2	&	188.9	&	141.8	&	461.1	\\
  \end{tabular*}
\end{table*}

In Fig.~\ref{fig:molecules}, we show, for the molecules in our set,
the performance measured as throughput of the SCF cycle and total
computational time as a function of the number of electrons.  As
expected, the computational time tends to increase with the number of
electrons, but there is a strong variation from system to system. This
variation is mainly explained by the physical size of each molecule,
that determines the size of the grid that is used in the
simulation. The number of iterations required for eigensolver and
self-consistency convergence can also change from one system to the
other, affecting the total calculation time. From
Fig.~\ref{fig:molecules}a is clear that as the size of the system
increases, the GPU becomes more efficient, with a maximum throughput
of 90 GFlops for the largest molecule tested, C\s{180}.

\begin{figure}
  \centering
 \includegraphics[width=\columnwidth]{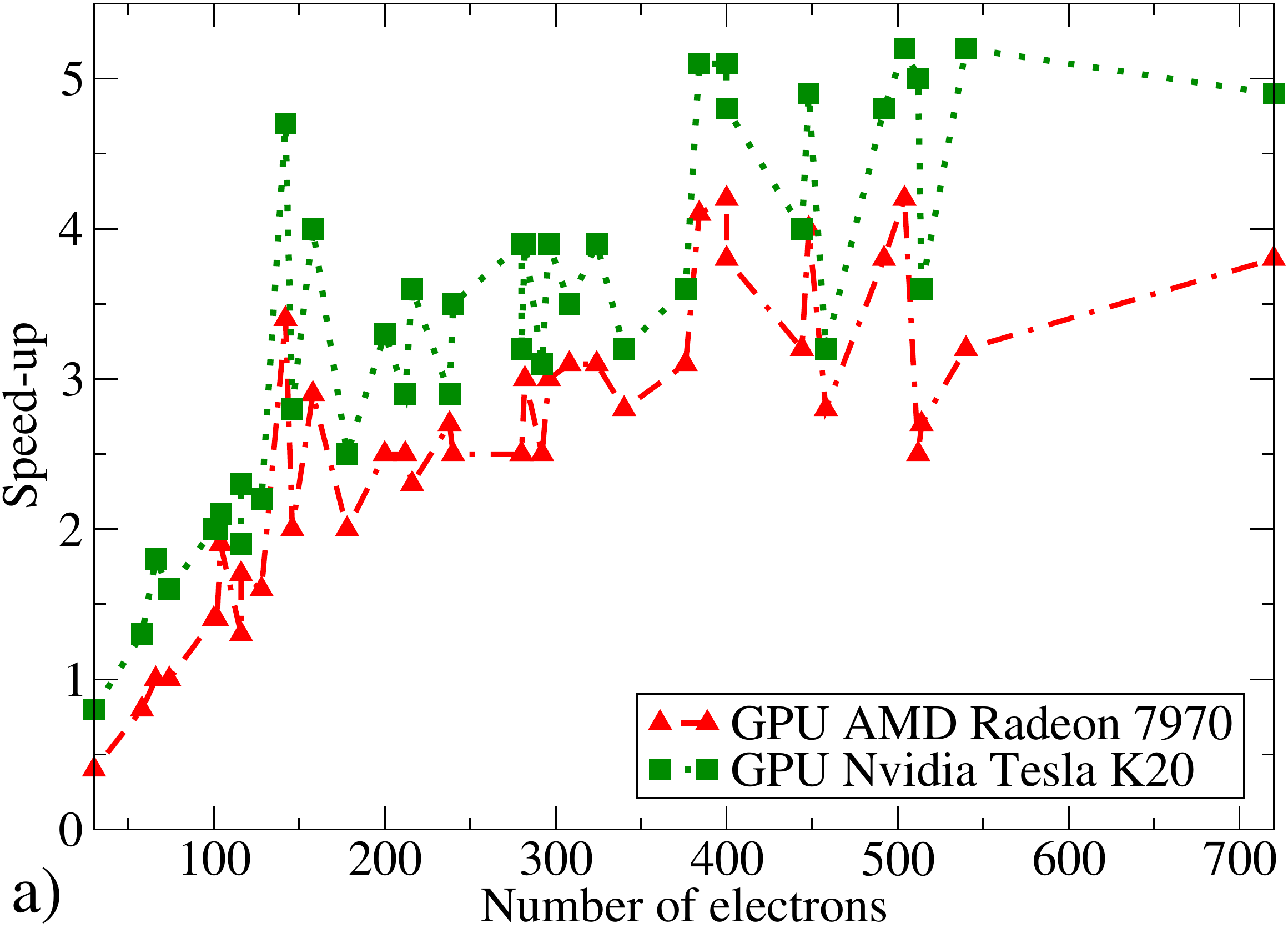}
 \includegraphics[width=\columnwidth]{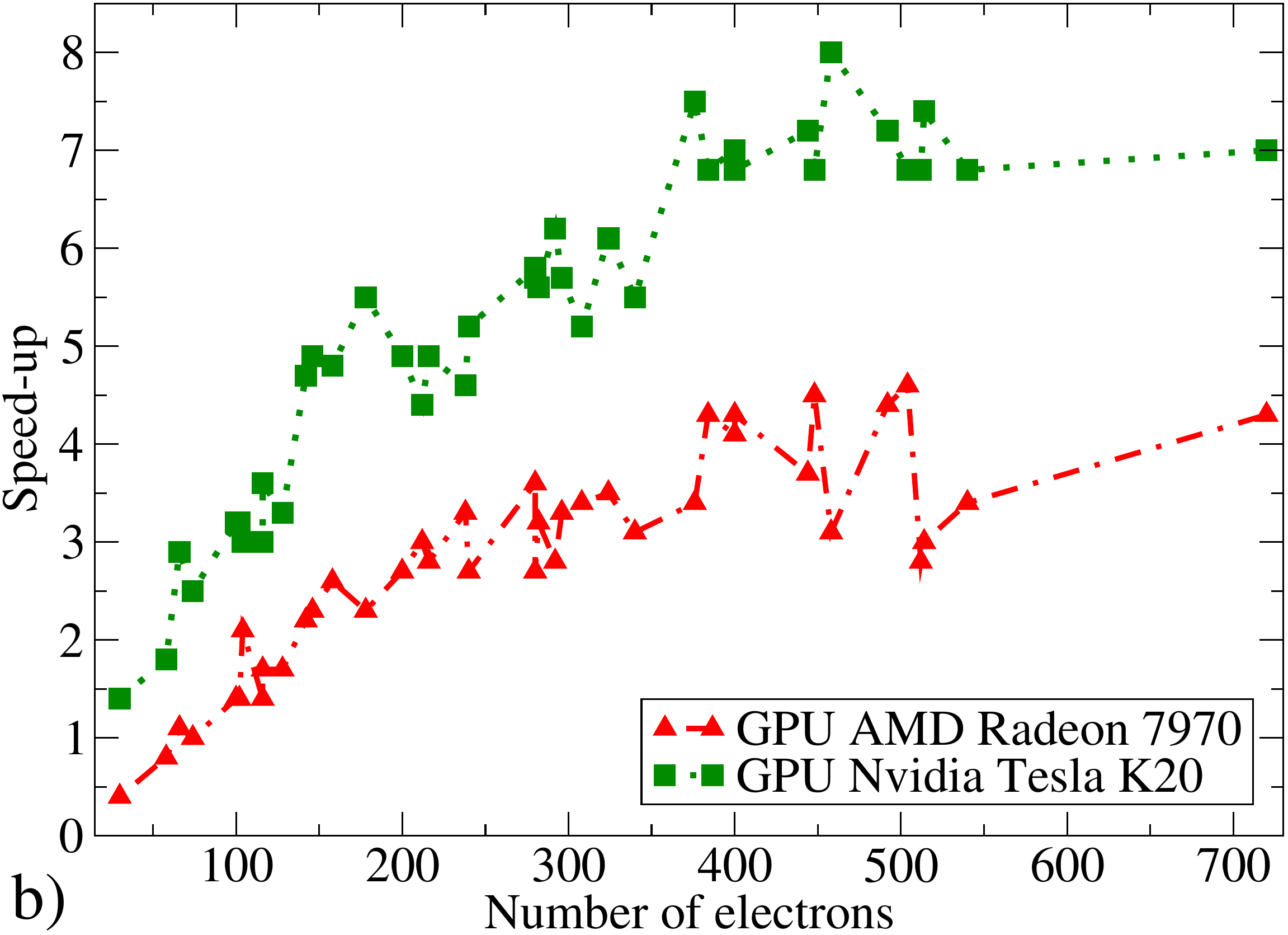}
 \caption{Speed-up of the GPU calculation with the respect to the CPU
   for different molecules as a function of the number of valence
   electrons. a) Speed-up calculated from the total calculation
   time. b) Speed-up computed from the time spent in the SCF-cycle
   (without considering initializations). The reference CPU is an
   Intel Core i7 3820 using 8 threads.}
  \label{fig:speedup}
\end{figure}

We now measure the speed-up of the GPUs with respect to the CPU
version. In Fig.~\ref{fig:speedup}a we plot the speed-up measured
using the total computational time. The maximum value we get is 5.2x
for the Nvidia GPU and 4.2x for the AMD GPU. If we only consider the
time spent in the SCF cycle and ignore the initialization time, the
speed-up is 8.0x for the Nvidia GPU and 5.2x for the AMD GPU, the
curve also becomes more regular, hinting that much of the variation in
the computational time for systems with similar number of electrons
comes from initialization routines.

While the speed-ups are not as large as some that have been reported
in the literature, there are several factors to consider when
analyzing GPU speed-ups. First of all, the maximum speed-up we could
obtain is given by the peak-performance ratio between the GPU and the
CPU, which is approximately 8x for and the AMD GPU and 10x for the
Nvidia card. If performance is limited by the memory bandwidth, then
the maximum speed-up is reduced to 5x (AMD) or 6x (Nvidia). The CPU
code taken as reference is also important. In this case we are
comparing code that uses the similar optimization strategies on the
CPU and the GPU, and in both the cases it has been parallelized to use
all the execution units available on each processor. This is not the
case, for example, when a full GPU is compared against a single core
of a CPU.


\subsection{Comparison with Terachem}

In order to make an exhaustive evaluation of the performance of our
approach, we compare it with another GPU-accelerated DFT
implementation, the \textsc{terachem} code~\cite{Ufimtsev2008a,
  Ufimtsev2008b, *Ufimtsev2009a, *Ufimtsev2009b, *Luehr2011}. \textsc{Terachem} uses Gaussian type orbitals (GTOs) as
a basis for the expansion of the molecular orbitals: the traditional
approach used in quantum chemistry. \textsc{Terachem} has been
extended to perform different types of simulations like excited
states~\cite{Isborn2011} or \textit{ab-initio} molecular
dynamics~\cite{Ufimtsev2011}, and thanks to the computational power
offered by GPUs, it has been used to study challenging systems like
large proteins~\cite{Kulik2012,*Isborn2012}.

Since \textsc{octopus} and \textsc{terachem} use very different
simulation techniques, we take great care in making a significant
comparison. The main issue is to select discretization parameters that
produce a similar level of approximation. We take as reference the
caffeine molecule, C\(_8\)H\(_{10}\)N\(_4\)O\(_2\) in the
Becke-Lee-Yang-Parr (BLYP) XC 
approximation~\cite{Becke1988,*Lee1988,*Miehlich1989}. In
\textsc{terachem} we select the \textit{6-311g*} basis that has an
error in the total energy of 5 \review{millihartree} per atom, with respect to a
calculation with the \textit{aug-cc-pvqz} basis. We then look for grid
parameters that give a similar error, this time taking as reference
the converged real-space result. The selected grid is a union of
spheres of radius 5.5 Bohr around each atom and a spacing of 0.41
Bohr. However, the real-space approach has an additional
approximation, as it requires pseudo-potentials so that the ionic
potential is smooth enough to be represented in a uniform grid. To
minimize the effect of this difference in computation time and to
compare the actual implementation, we test molecules composed mainly
of first and second-row elements.

\begin{figure}
  \centering
  \includegraphics[width=\columnwidth]{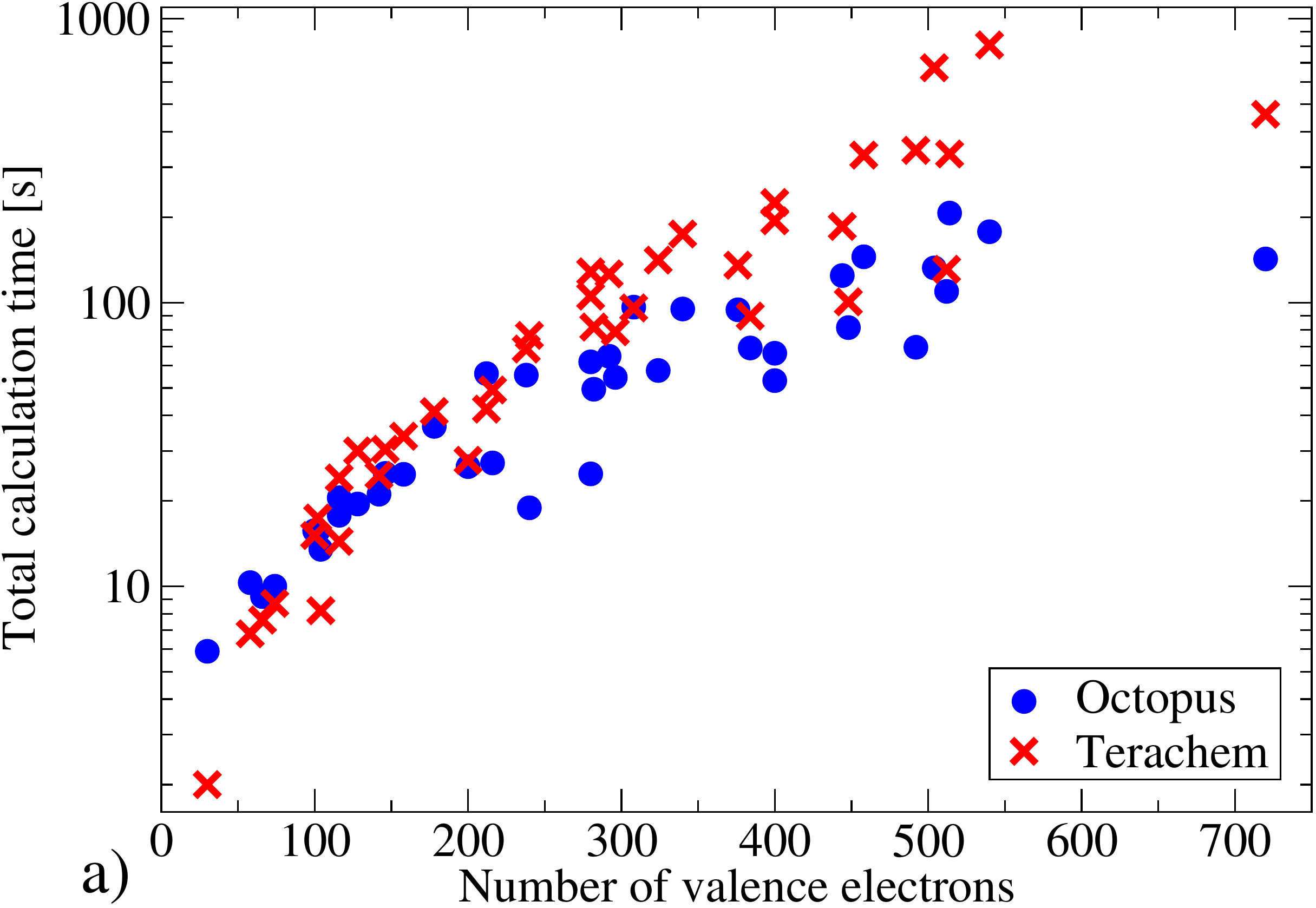}
  \includegraphics[width=\columnwidth]{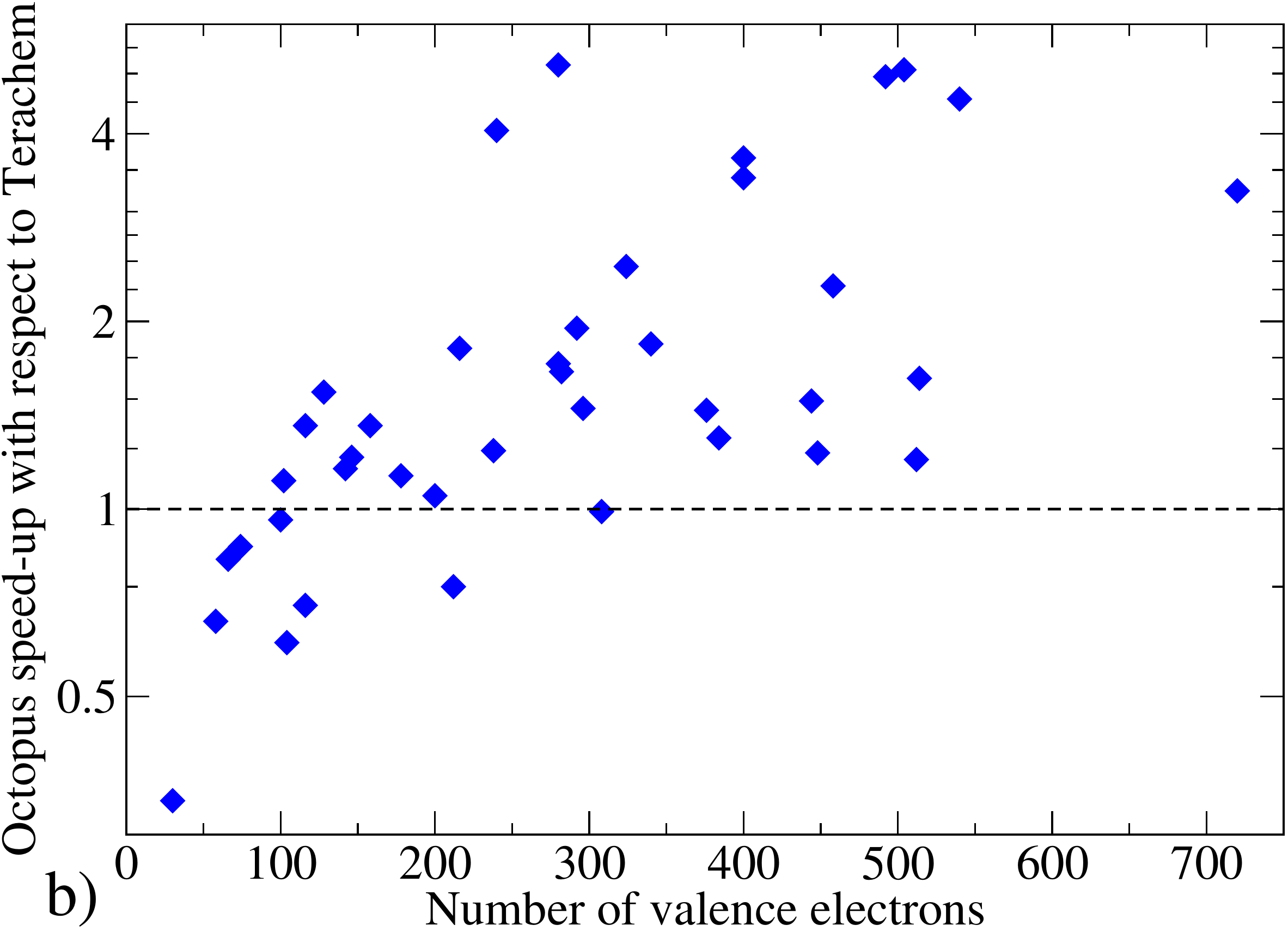}
  \caption{Numerical performance comparison between our GPU
    implementation (\textsc{octopus}) and the \textsc{terachem}
    code. a) Comparison of the total calculation time as a function of
    the number of valence electrons. b) Speed-up of our implementation
    with respect to \textsc{terachem} (run time of \textsc{terachem}
    divided by the run time of \textsc{octopus}). The calculations are
    single-point energy evaluations performed on a set of 40
    molecules, running on a Nvidia Tesla K20 GPU. The list of
    molecules and the calculation times are given in table
    \ref{tab:molecules}.}
  \label{fig:terachem}
\end{figure}

In Fig.~\ref{fig:terachem}, we compare the timings for both codes for
the same set of systems used in section~\ref{sec:performance}
(table~\ref{tab:molecules}). We show the comparison between absolute
times and also the relative performance between the two DFT
implementations. We can see that \textsc{terachem} tends to be faster
for smaller systems, while \textsc{octopus} has an advantage for
systems with more than 100 electrons. It is difficult to generalize
these results due to the different simulation approaches and their
different strengths and weaknesses. For example, our current
implementation will certainly be much slower than \textsc{terachem}
for hybrid HF-DFT XC approximations~\cite{Becke1993} due to the cost
of applying exact-exchange operator in real-space. However, we can
conclude that \review{for pure DFT calculations the
  real-space method can compete with the Gaussian approach}, and can
outperform it for some systems.

\section{Conclusions}

We have presented an approach for the implementation of real-space
density functional theory on GPUs. What we have shown is much more
than a re-implementation of the code in GPU language, but a scheme
designed to perform DFT calculations efficiently on massively parallel
processors.

Our approach is based on using blocks of KS orbitals as the basic data
object. This provides the GPU with enough data to perform efficiently,
something that would be harder to achieve by working on single
orbitals at a time. However, this approach is not applicable or does
not work efficiently for all operations, so in other cases a
block-of-points strategy is used. Many of these techniques are
applicable to other DFT discretization approaches, especially those
based on sparse representations like plane-waves or wavelets.

The efficiency of our approach is analyzed by examining several
parameters. We achieve a considerable throughput and speed-up with
respect to the CPU version of \textsc{octopus}. More importantly, in
comparison to a GPU-accelerated implementation of DFT based on
Gaussian basis sets, we find that calculation times are similar, with
our code being faster for several of the systems that were
tested. This is not to be taken lightly, as the GTO approach has been
designed and constantly improved with the specific purpose of
efficiently modeling molecular systems. The real-space method, on the
other hand, is a more general approach used to study different types
of partial differential equations.

We can conclude that the real-space formulation provides a good
framework for the implementation of DFT on GPUs, making real-space DFT
an interesting alternative for electronic structure calculations, as
it offers good performance, systematic control of the discretization
and the flexibility to study many classes of systems, including both
periodic and finite systems.

A particular advantage of real-space DFT is its potential for large
scale parallelization in distributed memory systems with tens of
thousands of
processors~\cite{Andrade2012a,Enkovaara2010,Hasegawa2011}. This is
something we want to apply in future work, by exploring the
combination of in-processor (OpenCL) and distributed memory (MPI)
parallelization for DFT calculations on GPU-based supercomputers.

\section{Computational methods}
\label{sec:methods}

Our numerical implementation is included in the {\sc Octopus} code
\cite{Marques2003,Castro2006,Andrade2012a} and it is publicly
available under the GPL free-software license~\cite{octopusdownload}. The
calculations were performed with the development version
(\textit{octopus superciliosus}, svn revision 10562). GPU support is
also available in the 4.1 release of Octopus. 

Since \textsc{octopus} is written in Fortran 95, we wrote a wrapper
library to call OpenCL from that language. This library is called
\textsc{FortranCL} and it is available as a standalone package under a
free-software license~\cite{FortranCL}.

All calculations were performed using the default pseudo-potentials of
\textsc{octopus}, filtered to remove high-frequency
components~\cite{Tafipolsky2006}. The grid for all simulation is a
union of spheres of radius 5.5 Bohr around each atom with a uniform
spacing of 0.41 Bohr.

The GTO calculations were done with \textsc{terachem} (version v1.5K)
with the \textit{6-311g*} basis and \texttt{dftgrid = 1}. All other
simulation parameters were kept in its default values. For all
calculations we used the BLYP XC functional~\cite{Becke1988,*Lee1988,*Miehlich1989}.

The system used for the tests has an Intel Core i7 3820 CPU, which has
4 cores running at 3.6 GHz that can execute 2 threads each. The CPU
has a quad-channel memory subsystem with 16 GiB of RAM running at 1600
MHz. The GPUs are a AMD Radeon HD 7970 with 3 GiB of RAM and Nvidia
Tesla K20c with 5 GiB (ECC is disabled\review{, as the other
  processors do not support ECC}). Both GPUs are connected to a PCIe
16x slot, the AMD card supports the PCIe 3 protocol while the Nvidia
card is limited to PCIe 2. \textsc{Octopus} was compiled with the GNU
compiler (gcc and gfortran, version 4.7.2) with AVX vectorization
enabled. \review{For finite-difference operations, CPU vectorization is
  implemented explicitly using compiler directives. We use the Intel
  MKL (version 10.3.6) implementation of \textsc{blas} and
  \textsc{lapack} that is optimized for AVX.} We use the OpenCL
implementation from the respective GPU vendor: the AMD OpenCL version
is 1084.4 (VM) and the Nvidia one is 310.32 (OpenCL is not used for
the CPU calculations). All tests are executed with 8 OpenMP threads.

Total and partial execution times were measured using the
\texttt{gettimeofday} call. The throughput is defined as the number of
floating point additions and multiplications per unit of time. The
number of operations for each procedure is counted by inspection of
the code. For \textsc{terachem} the total execution time is obtained
from the program output.



    



\begin{acknowledgments}
  We thank N.~Suberviola, J.~Muguerza and A.~Arruabarrena for useful
  discussions, and D.~Strubbe, J.~Alberdi, M.\,A.\,L.~Marques and all
  the \textsc{octopus} development team for their effort in developing
  and maintaining the code (in particular for detecting and fixing
  many of the bugs introduced while implementing the GPU support).

  We would like to acknowledge Nvidia for support via the Harvard CUDA
  Center of Excellence, and both Nvidia and Advanced Micro Devices
  (AMD) for providing the GPUs used in this work. This work was
  supported by the Defense Threat Reduction Agency under Contract No
  HDTRA1-10-1-0046.
\end{acknowledgments}



\bibliography{biblio}



\end{document}